\documentclass[journal]{IEEEtran}
\usepackage{graphicx}
\usepackage{amstext}
\usepackage{algorithm}
\usepackage{algorithmic}
\usepackage{mathrsfs}
\usepackage{amssymb}
\usepackage{amsmath}
\usepackage{bm}
\allowdisplaybreaks[4]
\usepackage{epstopdf}
\usepackage{multicol}
\usepackage{stfloats}
\usepackage{url}
\usepackage{color}
\usepackage{enumerate}
\usepackage{breqn}
\usepackage{bbm}
\usepackage{cite}
\usepackage{caption}
\usepackage{enumitem}
\usepackage{array}
\ifCLASSINFOpdf
\else
\fi
\begin{document}

\title{Blockchain for the Internet of Vehicles towards Intelligent Transportation Systems: A Survey}

\author{Muhammad Baqer Mollah,~\IEEEmembership{Member,~IEEE,}
        Jun Zhao,~\IEEEmembership{Member,~IEEE,}
        Dusit Niyato,~\IEEEmembership{Fellow,~IEEE,}
        \\Yong Liang Guan,~\IEEEmembership{Senior Member,~IEEE,}
        Chau Yuen,~\IEEEmembership{Senior Member,~IEEE,}
        Sumei Sun,~\IEEEmembership{Fellow,~IEEE,}
        \\Kwok-Yan Lam,~\IEEEmembership{Senior Member,~IEEE,}
        and Leong Hai Koh,~\IEEEmembership{Senior Member,~IEEE}%

\thanks{The work of J. Zhao is supported by 1) Nanyang Technological University (NTU) Startup Grant, 2) Alibaba-NTU Singapore Joint Research Institute (JRI), 3) Ministry of Education Singapore Academic Research Fund Tier 1 RG128/18, Tier 1 RG115/19, Tier 1 RT07/19, Tier 1 RT01/19,  RG24/20, and Tier 2 MOE2019-T2-1-176, 4) NTU-WASP Joint Project, 5) Singapore National Research Foundation (NRF) under its Strategic Capability Research Centres Funding Initiative: Strategic Centre for Research in Privacy-Preserving Technologies \& Systems (SCRIPTS), 6) Energy Research Institute @NTU (ERIAN), 7) Singapore NRF National Satellite of Excellence, Design Science and Technology for Secure Critical Infrastructure NSoE DeST-SCI2019-0012, 8) AI Singapore (AISG) 100 Experiments (100E) programme, and 9) NTU Project for Large Vertical Take-Off \& Landing (VTOL) Research Platform. The work of D. Niyato is supported by the National Research Foundation (NRF), Singapore, under Singapore Energy Market Authority (EMA), Energy Resilience, NRF2017EWT-EP003-041, Singapore NRF2015-NRF-ISF001-2277, Singapore NRF National Satellite of Excellence, Design Science and Technology for Secure Critical Infrastructure NSoE DeST-SCI2019-0007, A*STAR-NTU-SUTD Joint Research Grant on Artificial Intelligence for the Future of Manufacturing RGANS1906, Wallenberg AI, Autonomous Systems and Software Program and Nanyang Technological University (WASP/NTU) under grant M4082187 (4080), Singapore Ministry of Education (MOE) Tier 1 (RG16/20), and NTU-WeBank JRI (NWJ-2020-004), Alibaba Group through Alibaba Innovative Research (AIR) Program and Alibaba-NTU Singapore Joint Research Institute (JRI). The work of Y. L. Guan and C. Yuen is supported by A*STAR under its RIE2020 Advanced Manufacturing and Engineering (AME) Industry Alignment Fund – Pre Positioning (IAF-PP) (Grant No. A19D6a0053). Any opinions, findings and conclusions or recommendations expressed in this material are those of the author(s) and do not reflect the views of A*STAR. (Corresponding authors: \textit{Jun Zhao and Muhammad Baqer Mollah})}%
\thanks{M. B. Mollah, J. Zhao, D. Niyato, and K. Y. Lam are with the School of Computer Science and Engineering, Nanyang Technological University, Singapore 639798 (Email: muhd.baqer@ntu.edu.sg; junzhao@ntu.edu.sg; dniyato@ntu.edu.sg; kwokyan.lam@ntu.edu.sg).}%
\thanks{Y. L. Guan is with the School of Electrical and Electronic Engineering, Nanyang Technological University, Singapore 639798 (Email: eylguan@ntu.edu.sg).}%
\thanks{L. H. Koh is with the Energy Research Institute, Nanyang Technological University, Singapore 639798 (Email: lhkoh@ntu.edu.sg).}%
\thanks{C. Yuen is with the Engineering Product Development Pillar, Singapore University of Technology and Design, Singapore 119613 (Email: yuenchau@sutd.edu.sg).}%
\thanks{S. Sun is with the Institute for Infocomm Research (I2R), Agency for Science, Technology and Research (A*STAR), Singapore 138632. She is also with the Infocomm Technology Cluster, Singapore Institute of Technology, Singapore 138683 (Email: sunsm@i2r.a-star.edu.sg).}%
\thanks{Copyright (c) 20xx IEEE. Personal use of this material is permitted. However, permission to use this material for any other purposes must be obtained from the IEEE by sending a request to pubs-permissions@ieee.org.}
      }

\markboth{Accepted at IEEE INTERNET OF THINGS JOURNAL 2020}%
{Shell \MakeLowercase{\textit{et al.}}: Bare Demo of IEEEtran.cls for IEEE Journals}

\maketitle

\begin{abstract}
Internet of Vehicles (IoV) is an emerging concept that is believed to help realise the vision of intelligent transportation systems (ITS). IoV has become an important research area of impactful applications in recent years due to the rapid advancements in vehicular technologies, high throughput satellite communication, Internet of Things and cyber-physical systems. IoV enables the integration of smart vehicles with the Internet and system components attributing to their environment such as public infrastructures, sensors, computing nodes, pedestrians and other vehicles. By allowing the development of a common information exchange platform between vehicles and heterogeneous vehicular networks, this integration aims to create a better environment and public space to the people as well as to enhance safety for all road users. Being a participatory data exchange and storage, the underlying information exchange platform of IoV needs to be secure, transparent and immutable in order to achieve the intended objectives of ITS. In this connection, the adoption of blockchain as a system platform for supporting the information exchange needs of IoV has been explored. Due to their decentralized and immutable nature, IoV applications enabled by blockchain are believed to have a number of desirable properties such as decentralization, security, transparency, immutability, and automation. In this paper, we present a contemporary survey on the latest advancement in blockchain for IoV. Particularly, we highlight the different application scenarios of IoV after carefully reviewing the recent literatures. We also investigate several key challenges where blockchain is applied in IoV. Furthermore, we present the future opportunities and explore further research directions of IoV as a key enabler of ITS.
\end{abstract}

\begin{IEEEkeywords}
Blockchain, Connected Vehicles, Internet of Things, Internet of Vehicles, Intelligent Transportation System, Security, Smart Vehicle, Smart Transportation.
\end{IEEEkeywords}

\IEEEpeerreviewmaketitle

\section{Introduction}
\IEEEPARstart{B}{lockchain} was initially introduced as a distributed ledger of the Bitcoin \cite{nakamoto2019bitcoin, tschorsch2016bitcoin, belotti2019vademecum}  system for the purpose of addressing the double-spending problem of the cryptocurrency. One of the key features of blockchain is that, thanks to the immutability of the distributed ledger, it allows transacting parties and stakeholders to establish trust among untrusted entities in a decentralized manner. Due to the huge success of Bitcoin, blockchain has attracted great attention of the research community as an emerging technology. Although blockchain originated as an infrastructure for cryptocurrency, it has become a distributed system technology that inspired and drove a wave of paradigm shift from centralized to decentralized and dynamic system architecture. Blockchain-based architecture is decentralized and open as it is implemented by a number of distributed nodes, each of which contain a replica of the cryptographically-chained bitcoin transaction records, organized in blocks, agreed upon by some consensus protocols among the blockchain nodes. The cryptographic chaining of the blocks, together with the distributed consensus protocol, ensure the immutability of blockchain. The opennes and immutability of blockchain allow anyone to verify the history of bitcoin transactions, hence defeating any attempt to tamper the transaction history without being detected. As such, trust among untrusted entities can be established in a decentralized manner. That means, once appended to the chain, the records cannot be modified. The decentralized, open, and unmodified nature makes the blockchain a transparent, fully distributed, and publicly verifiable architecture as well. Furthermore, as the records are replicated across the multiple distributed nodes, the blockchain architecture inherently addressed the single point of failure problem. As a result, a wide range of applications are exploring to adopt blockchain technology by taking benefits of the aforesaid attractive properties.

On the other hand, with increasing advancements of vehicular technologies, satellite communications \cite{li2019spectral, li2017joint} and cyber-physical systems (CPS), vehicles are becoming smarter and more autonomous or semi-autonomous than before \cite{hussain2018autonomous, kazmi2019infotainment, sun2017secure}. The Internet of Vehicles (IoV) concept \cite{tanwar2019tactile, sharma2019survey, dai2019artificial} has emerged to support the realization of future intelligent transportation system (ITS), in which a growing number of smart vehicles are interconnected with the Internet. It is envisioned that ITS will become a reality in one decade through the adoption of IoV. To date, a number of initiatives have been launched to encourage development of ITS, particularly into smart cities, such as ERTICO - ITS Europe \cite{misc2} and CityVerve Manchester \cite{misc3}. Besides, IoV also enables the seamless interconnection among smart vehicles, roadside infrastructures and pedestrians in order to meet the evolving functional requirements of ITS and to facilitate the vehicle-to-everything (V2X) paradigm. Moreover, cloud computing, edge computing, artificial intelligence, and modern vehicular communication technologies are accelerating the evolution of IoV. The ultimate aim of IoV is to make a common, seamless and connected platform to exchange data and resources for smart vehicles. The capabilities of these data and resources exchanges are driving a number of vehicular applications such as enhanced road safety, driving safety, traffic efficiency, smart parking and entertainment services.

With the rapid development of vehicular applications \& services, the growing number of smart vehicles are expected to produce and exchange an enormous amount of data, and the network traffic to be managed will be significantly huge. At the same time, the high mobility, low latency, context complexity, and heterogeneity characteristics of IoV will also face substantial difficulties when utilizing traditional cloud-based storage and management directly. Additionally, it is also difficult to ensure strong interoperability and compatibility among IoV entities belonging to different service providers. Hence, the data exchange and storage platform for IoV need to be decentralized, distributed, interoperable, flexible and scalable in order to cater for the future growth of IoV and fully realize the potential of ITS. Furthermore, being distributed and decentralization, the platform is inherently more vulnerable to cyber attacks, hence it is essential to ensure security, privacy, and trust of the IoV data. Consequently, the blockchain technology along with modern cryptographic techniques, and edge computing, have presented great opportunities in a number of IoV applications already. 

\textit{Contributions of this Survey:}
In this survey, we focus on the role blockchain in IoV, and investigate existing research works that have been presented in different literature. The contributions of this survey are summarized below.

\begin{itemize}
    \item We present a comprehensive survey of the techniques for integration of blockchain and IoV paradigm towards building a future ITS, starting with describing the preliminary background including blockchain technology, edge computing, ITS and IoV.
    \item We discuss the challenges associated with IoV, highlight the importance as well as motivations of the convergence of blockchain and IoV, and point out the specific challenges which can be addressed by blockchain.
    \item We present the state-of-art research efforts and in-depth discussion on the adoption of blockchain for IoV scenarios, with a particular focus on vehicular data security, vehicle management, and on-demand transportation services.
    \item We highlight a number of blockchain-empowered IoV architectures, including the potential integration of blockchain, edge computing, vehicular communication systems, automotive technologies, and privacy preserving techniques.
    \item We identify and investigate the key challenges associated with blockchain integration with IoV, including security \& privacy, performance, IoV-specific \& optimized consensus, and incentive mechanisms.
    \item We outline a number of open issues and challenges as future research opportunities in the area of blockchain, IoV, and vehicular cyber-physical systems.
\end{itemize}

\textit{Related Surveys:} A number of recent surveys have presented some specific aspects of blockchain in certain IoT and CPS domains. For example, a literature survey on blockchain-assisted IoT system with a focus on privacy-preserving techniques and associated research issues is presented in \cite{hassan2019privacy}. Meanwhile, the approaches, opportunities, and challenges of the blockchain-enabled typical IoT and CPS are studied in \cite{makhdoom2019blockchain, cao2019internet, viriyasitavat2019blockchain, dai2019blockchain, koshy2020sliding, tseng2020blockchain, lockl2020toward, sharma2020blockchain, rathore2020survey, mehta2020blockchain}. The applications, benefits, and existing research advancements of the adoption of blockchain in Industrial IoT (IIoT) and Industry 4.0 are discussed in \cite{alladi2019blockchain, shen2020blockchain, luo2020blockchain, li2020blockchain, leng2020blockchain, sengupta2020comprehensive}. In order to improve the performance of cooperative robotics and its distributed control strategies, the architecture, definitions, and applications are presented in \cite{lopes2018overview, du2020overview}. The taxonomy and comparative analysis of latest solutions, architectures, and requirements of blockchain with application to smart cities are discussed in \cite{xie2019survey, hakak2020securing, nagel2020smart, zhang2020ldc, aujla2020blocksdn, bhushan2020blockchain}. A substantial effort has been highlighted to converge blockchain with smart health system in \cite{wang2019distributed, farouk2020blockchain, khan2020blockchain, de2020survey, chukwu2020systematic, hasselgren2020blockchain}. A comprehensive survey of recent developments in blockchain for future smart grid scenarios including advanced metering infrastructure (AMI), electric vehicles (EVs), EVs charging units management, energy CPS, distributed energy resources (DERs), and recent industrial initiatives is presented in \cite{mollah2019blockchain}. Also, in \cite{kuzlu2020realizing, aderibole2020blockchain, zhuang2020blockchain, alcaraz2020blockchain, miglani2020blockchain, dimobi2020transactive, hassan2019blockchain, bao2020survey, lu2019blockchain}, the blockchain roles specifically in smart grid as well as other energy systems (oil and gas) are discussed. The blockchain in 5G and beyond 5G/6G enabling technologies including edge and cloud computing, network function virtualization software defined networking, network slicing, device to device communications, and spectrum management is presented in \cite{nguyen2019blockchain, chaer2019blockchain, weiss2019application, liu2020blockchain, hewa2020role}. A number of works on the integration of blockchain and machine learning (ML) are presented in \cite{singh2020convergence, mohanta2020survey, gupta2020smart, singh2020blockiotintelligence, tanwar2019machine}. In Table~\ref{tab: summaryotherdomains}, a summary of the aforementioned blockchain applications is presented. However, to date, all previous surveys lack a survey of blockchain in the specific context of IoV. Hence, our work is distinct from all aforementioned surveys since we significantly present the roles of blockchain in IoV application scenarios. Moreover, a survey on blockchain in IoV will open the door for those who are going to apply the blockchain technology into the area of intelligent transportation system and its related research fields.

\begin{table}[ht]
    \centering
    \caption{Summary of the applications of blockchain in other emerging domains}
    \label{tab: summaryotherdomains}
\begin{tabular}{l|l}
\hline \hline
Application Domains & References \\
\hline
Internet of Things (IoT) and Cyber-Physical Systems (CPS) & \cite{hassan2019privacy, makhdoom2019blockchain, cao2019internet, viriyasitavat2019blockchain, dai2019blockchain, koshy2020sliding, tseng2020blockchain, lockl2020toward, sharma2020blockchain, rathore2020survey, mehta2020blockchain} \\
\hline
Industrial IoT (IIoT) and Industry 4.0 & \cite{alladi2019blockchain, shen2020blockchain, luo2020blockchain, li2020blockchain, leng2020blockchain, sengupta2020comprehensive} \\
\hline
Cooperative Robotics & \cite{lopes2018overview, du2020overview} \\
\hline
Smart city applications & \cite{xie2019survey, hakak2020securing, nagel2020smart, zhang2020ldc, aujla2020blocksdn, bhushan2020blockchain} \\
\hline
Smart health system & \cite{wang2019distributed, farouk2020blockchain, khan2020blockchain, de2020survey, chukwu2020systematic, hasselgren2020blockchain} \\
\hline
Smart grid and other energy domains & \cite{mollah2019blockchain, kuzlu2020realizing, aderibole2020blockchain, zhuang2020blockchain, alcaraz2020blockchain, miglani2020blockchain, dimobi2020transactive, hassan2019blockchain, bao2020survey, lu2019blockchain} \\
\hline
5G and beyond 5G/6G-enabling technologies & 
\cite{nguyen2019blockchain, chaer2019blockchain, weiss2019application, liu2020blockchain, hewa2020role} \\
\hline
Machine Learning & \cite{singh2020convergence, mohanta2020survey, gupta2020smart, singh2020blockiotintelligence, tanwar2019machine} \\
\hline\end{tabular}
\end{table}	

\begin{figure*} [h]
	\includegraphics[width=\linewidth]{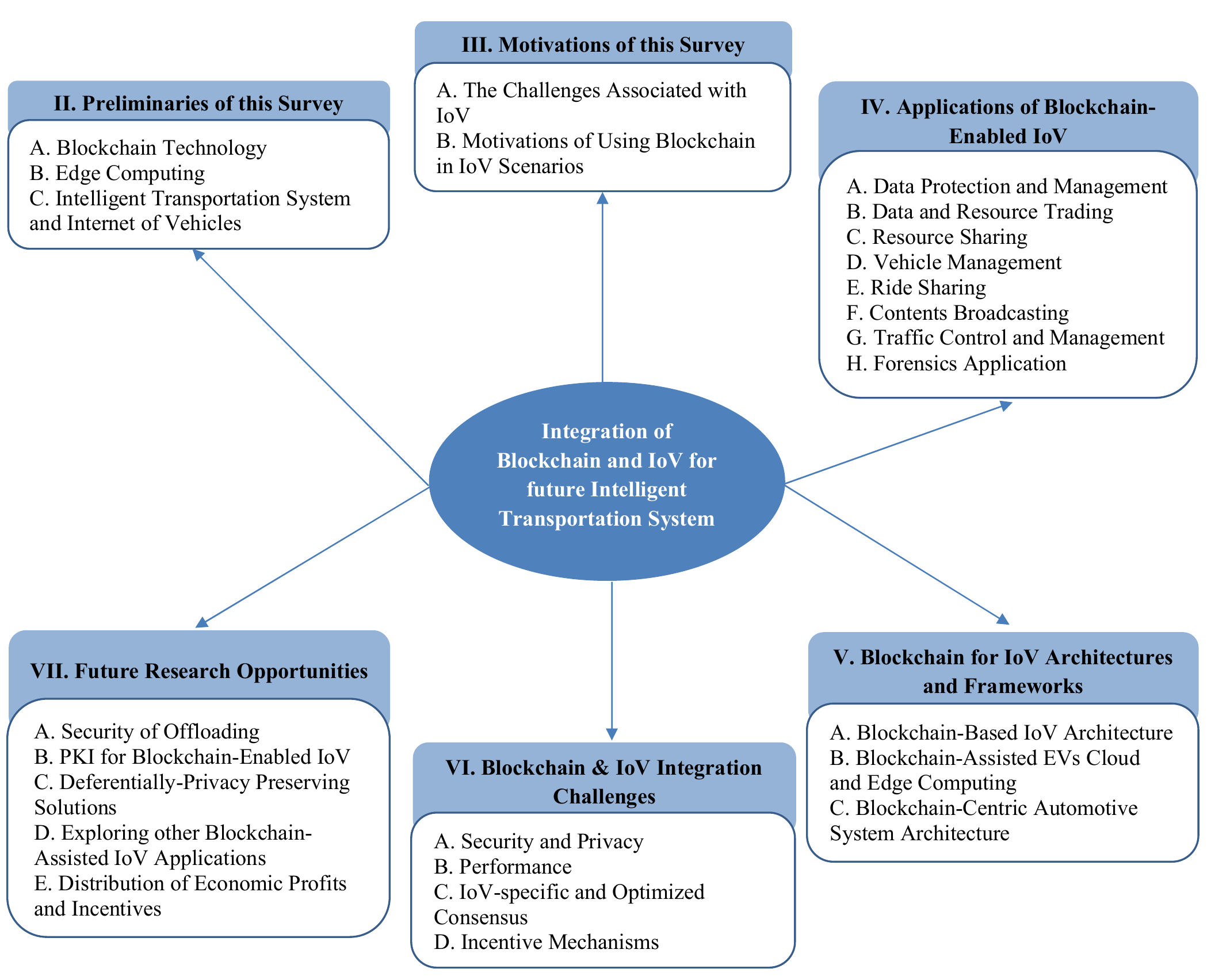}
	\caption{Outline of the key topics of this paper.}
    \label{fig: outlineofpaper}
\end{figure*}

\textit{Paper Organization:}
The rest of the paper is structured as follows. In Section II, we outline the preliminaries of this survey, which include an overview of blockchain technology, edge computing, intelligent transportation system (ITS), and the Internet of Vehicles (IoV). In Section III, we look into the challenges associated with IoV, and then, we highlight the motivations of adopting blockchain into IoV domain. In Section IV, we provide a comprehensive review and in-depth investigation of the recent works on blockchain-assisted IoV applications. In Section V, we present a discussion on four blockchain and IoV integration challenges. In Section VI, we explore the architectures and frameworks which employed the blockchain and IoV. In Section VII, we point out the potential future research opportunities. Finally, in Section VIII, we present the concluding remarks of this paper. As such, the key topics presented in this survey are outlined in Fig. \ref{fig: outlineofpaper}. Table \ref{tab: acronyms} highlights a list of major acronyms along with their definitions used in this paper.

\begin{table}[ht]
    \centering
    \caption{Summary of Acronyms.}
    \label{tab: acronyms}
\begin{tabular}{l|l}
\hline \hline
Acronym                                      & Their Definition \\
\hline
IoT	 									 & Internet of Things \\
\hline
IoV                         					 & Internet of Vehicles \\
\hline
ITS										     & Intelligent Transportation System \\
\hline
CPS									         & Cyber-Physical System \\
\hline
AV									         & Autonomous Vehicle \\
\hline
CAV										 & Connected and Autonomous Vehicle \\
\hline
ECU                      					 & Electronic Control Units \\
\hline
OBU											 & Onboard Units \\
\hline
CAN											 & Controller Area Network \\
\hline
EDR											 & Event Data Recorders \\
\hline
LIN									     & Local Interconnect Network \\
\hline
VCS										     & Vehicular Communication Systems \\
\hline
BSM										     & Basic Safety Message \\
\hline
V2X											 & Vehicle-to-Everything \\
\hline
V2V										     & Vehicle-to-Vehicle \\
\hline
V2I											 & Vehicle-to-Infrastructure \\
\hline
V2G	 										 & Vehicle-to-Grid \\
\hline
RSU											 & Roadside Units \\
\hline
AI											 & Artificial Intelligence \\
\hline
PUF	 										 & Physical Unclonable Function \\
\hline
PKI							  		     & Public Key Infrastructure \\
\hline
CA									         & Certificate Authority \\
\hline
DSRC									         & Dedicated Short Range Communication \\
\hline
ECDSA									         & Elliptic Curve Digital Signature Algorithm \\
\hline
EVs									         & Electric Vehicles \\
\hline
\end{tabular}
\end{table}	

\section{Preliminaries of this Survey}
In this section, we present a brief discussion on the preliminaries of this survey, in where we present the overviews and recent advances. The preliminaries include blockchain technology, edge computing, and Intelligent Transportation System \& Internet of Vehicles.

\subsection{Blockchain Technology}
Blockchain is a collection of blocks, where the blocks stores transactions, records, and scripts, and all blocks are linked together to build a chain based on some cryptographic techniques. The newly generated blocks are continuously affixed to the chain in a digital ledger, and the ledger is maintained by all participants in the network. Hence, blockchain is also called a distributed ledger technology (DLT). As mentioned in introduction, blockchain enables a platform to do trusted tasks and transactions in an untrusted environment without requiring a trusted entity. Generally, the blockchain technology relies on four key techniques such as cryptography, consensus, structure, and smart contract. A short discussion of these key techniques is given below.

\textit{Consensus Mechanism:} Since blockchain technology does not reply on any trusted entity, consensus mechanisms are used to establish trust among the untrusted entities. The aim of these mechanisms is to enable entities to agree on a single version of valid block to ensure a transparent and consistent outlook, which ultimately solves forks and conflicts within the network. A number of consensus mechanisms have been proposed and implemented in different blockchain applications. Each mechanism has unique rules and algorithms which form the requirement to be followed by the entities/nodes to include new blocks to the chain. Some of the notable consensus mechanisms are Proof of Work (PoW) \cite{nakamoto2019bitcoin}, Proof of Stake (PoS) \cite{misc1}, and Practical Byzantine Fault Tolerance (PBFT) \cite{castro1999practical}. 

\textit{Cryptography:} Cryptographic techniques are adopted in blockchain technology to ensure the security, privacy, and anonymity. For example, the applications of cryptographic hash function are quite common in blockchain. Each block in blockchain contains the hash value of previous block’s transactions or records which connects each block to the previous block in order to build a chain. Once the blocks connect together, they will become immutable as the hash values are unique. Moreover, another application of hash is Merkle hash tree \cite{merkle1987digital}, which is used to make a summary of all the previous transactions or records in a block. The Merkle hash tree is able to decrease the storage overhead, and lightweight entities can use it to check the integrity quickly by it. In this way, the hash values ensure integrity, and make the blockchain tamper-proof.

In addition to the hash, another most commonly used cryptographic technique in blockchain is the digital signature. The usage of the digital signature is to guarantee authentication as well as non-repudiation. The Elliptic Curve Digital Signature Algorithm (ECDSA) scheme is widely applied in many blockchain applications like Bitcoin. Moreover, ring signature and multi-signature schemes are also used in some blockchains to hide the signer’s identity. Besides the hash function and digital signature, to improve the privacy of the block contents, the zero-knowledge proof is utilized in blockchain.

\textit{Smart Contract:} The smart contract are scripts resided in blocks of blockchain. The scripts are able to execute automatically once triggered or some pre-defined rules are met. The smart contract has become popular recently after the introduction of blockchain. The main aim of implementing the smart contract in blockchain is to develop a highly autonomous system which can provide efficient and consistent services without depending on any trusted entity.

\textit{Data Structure and Management:} Blockchain became successful in Bitcoin and popular among other applications due to its unique data structure. Although the linear structure is quite common, another type of structure is also available which is called DAG (Directed Acyclic Graph). Unlike linear structure, blocks are connected in a DAG with their previous multiple blocks. Apart from data structure, there are three types of data management techniques for blockchain, such as on-chain, off-chain, and side-chain. In on-chain, the block contents are recorded in blockchain which is visible to all entities of the network, whereas in off-chain, some contents are recorded and processed outside of the network for quick execution usually by trusted entity. On the other hand, side-chain \cite{back2014enabling}, is completely independent blockchain which runs in parallel to the main chain and maintains a link to the main chain. It allows users to transfer and use their cryptocurrencies or assets to this supplementary chain, and also, return back to the main chain.

\subsection{Edge Computing}
The edge computing concept is introduced as an extension of cloud computing to bring its capability to the edge of the network. Different literatures use different names of edge computing, such as fog computing and cloudlets, but the main aims of this concept are almost similar. 

Instead of depending on the cloud, the computational processing and data storage in edge computing are performed near the data source to handle the data locally which is collected from the user devices. Edge computing offers location-aware, low-latency, and real-time applications and services. It can also save bandwidth of transferring data to the cloud computing node which is located remotely. Although edge computing brings a number of benefits, the distributed nature of it introduces security and privacy challenges.

Edge computing has become an integral part of IoT and CPS applications \cite{ni2019toward, sodhro2019artificial, khan2019edge, liang2020towards, li2019advances, song2016cyber}, IoV applications \cite{ji2020artificial, zhang2019mobile, ning2019mobile, dai2020deep}, and blockchain \cite{xiong2018mobile, gai2019permissioned, guo2020blockchain, yang2019integrated}, to support a massive number of smart devices by running computational-intensive tasks and storage data at the edge. Specifically, in the context of IoV and ITS, to improve the quality of services while considering the mobility, the edge computing nodes are distributed frequently alongside the roads.

\subsection{Intelligent Transportation System and Internet of Vehicles}
With the advancements of communications, sensing, and electronic systems, the conventional embedded systems as well as controllers are being replaced by an advanced type of system which is referred as CPS. This CPS is usually connected to the Internet technologies to make a bridge between the cyber and physical worlds.

Recently, CPS has become popular and has been widely implemented from all aspects of our lives to industries. The ITS, considered as future transportations, is one of the examples of CPS. The ultimate goal of ITS is to develop more comfortable, safer, dynamic, and efficient transportation as well as urban infrastructures.

\begin{figure*} [b]
    \centering
	\includegraphics[width=\linewidth]{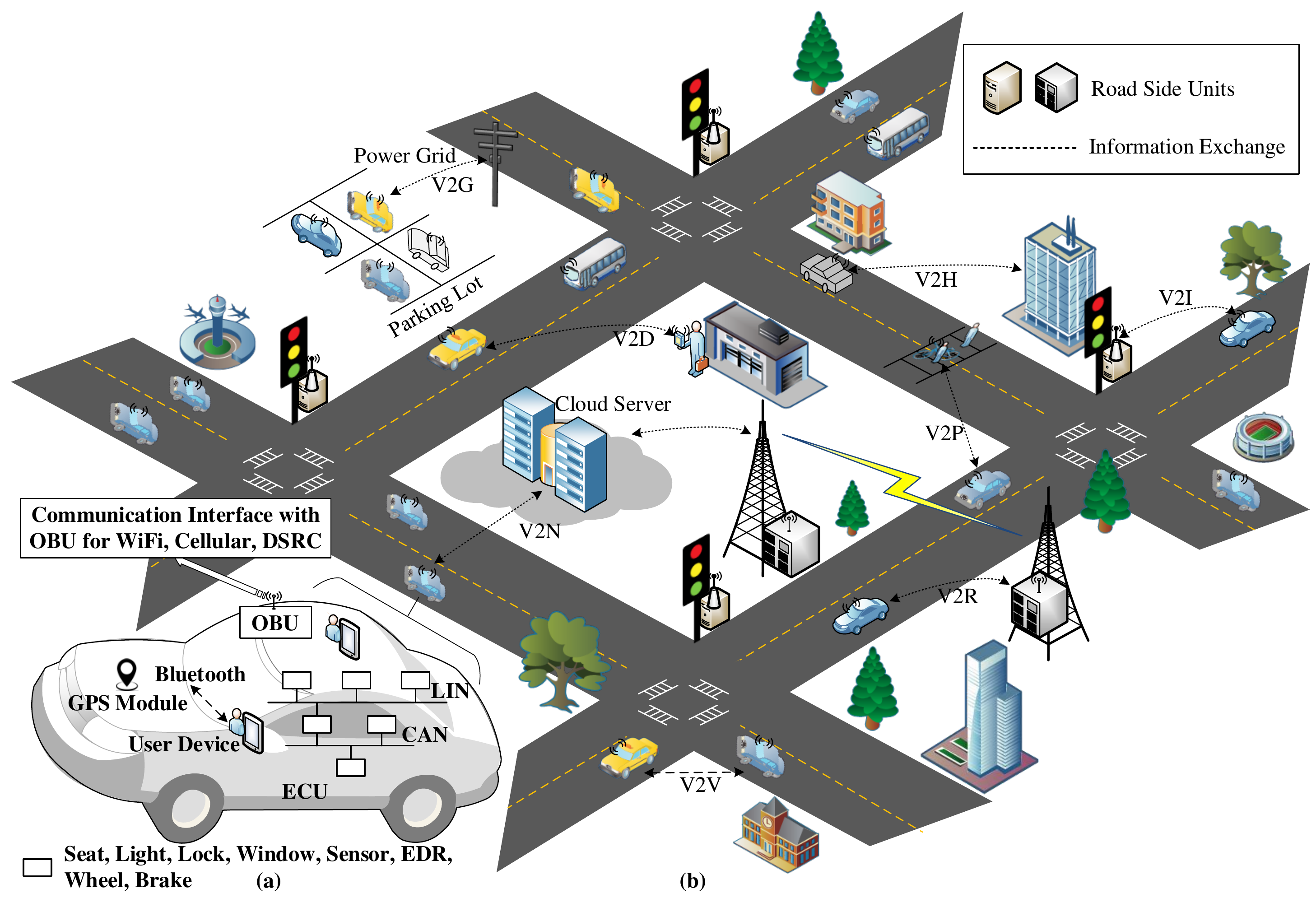}
	 \caption{(a) An overview of the communication structure of a smart vehicle; (b) The basic vehicular communication framework of ITS mainly contains vehicles, road-side units, smart devices, pedestrians, infrastructure, people, homes, grids as well as five types of V2X communications such as vehicle-to-road-side unit (V2R), vehicle-to-infrastructure (V2I), vehicle-to-vehicle (V2V), vehicle-to-pedestrians (V2P), vehicle-to-devices (V2D), vehicle-to-networks (V2N), vehicle-to-grid (V2G), and vehicle-to-home (V2H)}
	 \label{fig: overallits}
\end{figure*}

At the same time, the automobile industries are developing the technologies behind the smart vehicles, and thus, the vehicles are changing the experience and way of travelling. The smart vehicles as shown in in Fig. \ref{fig: overallits} (a) are typically equipped with in-vehicle computational \& storage units, EDR, control units such as ECUs \& OBUs, software \& firmware systems, diverse number of sensors, and multiple wireless devices. In this context, to take necessary actions, the control units rely on the data generated from the sensors and cameras, and the communications among all these components are built on different types of wired (CAN bus and LIN bus) and wireless (Bluetooth) technologies. These advanced technologies might incorporate into traditional vehicles soon. Moreover, with these advanced technologies, the vehicles are becoming even autonomous and semi-autonomous, which have the potential to create a revolution in the ITS.

A new paradigm called IoV is introduced within the ITS which is driven by the smart vehicles, Internet of Things (IoT), and AI techniques. In this paradigm, the vehicles are connected with each other, people, and infrastructures through communication technologies so that the vehicles are able to drive safely and intelligently through monitoring and sensing the neighboring environments. Hence, the IoV ecosystem is considered nowadays as an extended component of CPS. 

Besides the smart vehicles, the IoV ecosystem in ITS also consists of vehicular communication system, road-side units (RSU), and cloud platforms as shown in Fig. \ref{fig: overallits} (b). Here, the vehicular communication system is often referred to as vehicle to everything (V2X) which is nothing but a communication paradigm. V2X enables the communications from one vehicle to any nodes (i.e., vehicle, infrastructure, grid, pedestrian etc.), and vice versa. On the other hand, the RSUs usually act as access points and base stations to support VCS, and also, as edge computing nodes if they have computational and storage capabilities.

\section{Motivations of this Survey}

In this section, we first highlight the challenges associated with the IoV paradigm. Then, we present the motivations of blockchain and IoV integration to show which promising opportunities will be brought by blockchain for IoV.

\subsection{The Challenges Associated with IoV}
The developments in the IoV field are more rapid than ever before, due to the advancements of Internet technologies and vehicle equipment. Basically, this promising IoV field is envisioned to enrich the vehicular services, transportation infrastructures, and people’s lifestyle, at the same time, enhance drivers' safety in the near future. A massive amount of data will be introduced and outsourced to the cloud and edge storages from the vehicles as well as the vehicular services which will be innovated for IoV. The future vehicles will also have very good computational and storage resources. In order to offer a wide range of application services, these data and resources will be shared each other. Particularly, for artificial intelligence (AI) related applications, the vehicles reply on cloud and edge computing nodes to offload the tasks, and/or process tasks locally by sharing the resources with each other to reduce the latency and bandwidth.

However, the integration with existing Internet technologies to support IoV paradigm opens up many challenges \cite{chattopadhyay2020autonomous, hahn2019security, li2015art}, including security, privacy, trust, transparency, connectivity, and performance. The challenges associated with it will be increased with the growth of IoV connectivity. In this regard, many of challenges are inter-related with ITS. 

In fact, the IoV ecosystem poses a number of characteristics, and more specifically, among these some are unique compared to other IoT applications. Thus, IoV ecosystem might bring a number of novel challenges. In the following, such unique aspects of IoV ecosystem are described elaborately. 

\textit{High Mobility:} Unlike other IoT smart devices, in IoV scenarios, both driver-controlled and autonomous vehicles are considered as highly moving objects which usually run along the roads. Similarly, the running speeds of the vehicles may vary from one another which introduce diverse mobility particularly for manually driving vehicles. Thus, though vehicles have sufficient energy to deploy computational and communication resources, when the vehicles will connect with numerous peers, it will be difficult for vehicles to maintain a stable communication due to the diverse and high mobility regardless of dedicated channels. Specifically, high mobility characteristics possibly bring other challenges. 

\textit{Complexity in Wireless Networks:} The IoV ecosystem is relied on heterogeneous wireless communication network, where a number of wireless technologies coexist. And, in this ecosystem, the vehicles are connected with nearby vehicles, people as well as fixed road-side units through the wireless network. Typical technologies include Bluetooth, mmWave, Dedicated Short Range Communication (DSRC) etc., which enable various wireless network related services. For instance, Bluetooth and mmWave are able to provide the coverage less than 100m and 10m, respectively. Contrary of these two, DSRC has usually high communication coverage. Moreover, the vehicles change their network topologies while moving. Consequently, the impacts of complexity of networks on the IoV scenarios are significant. 

\textit{Latency-Critical Applications:} In many IoV applications, more efficient network protocols are required to exchange information to the nearby peers instead of far-away centralized cloud nodes. Indeed, those applications are often latency-sensitive and generally, have short to medium propagation distances. Thus, the maximum delay from source to destination should be low as much as possible for them. For example, emergency and safety related vehicular applications, where it is highly expected that the communication has to be done within a stipulated time limit in order to eliminate the possible unexpected situations like accidents. With this stringent delay constraint, the potential Internet-assisted technologies to be added in this IoV paradigm should be free from unnecessary communication delays in the Internet transmission. 

\textit{Scalability and Heterogeneity:} The vehicles which normally visit to a wide geographical area have a potentially convenient option to obtain the scalability through the road-side edge computing nodes, vehicular ad hoc networks, and wirelessly connected Internet technologies. Furthermore, the IoV components, which exhibit heterogeneous devices, protocols, and platforms, expect seamless integration with state-of the-art information and communication technologies by considering the heterogeneity. Apart from this, this heterogeneity of IoV components may add one more challenge which is difficulty to achieve interoperability. Indeed, interoperability refers the capability of IoV components to collaborate with each other in terms of information usages and exchange the information among the sectors, centers, and systems including both software and hardware. 

\textit{Artificial Intelligence (AI):} The application of AI techniques is going to become a part and parcel to assist the vehicles and vehicle users in many IoV application scenarios to cope with the utilization of a wide range of vehicular data. The whole procedures from data collection to deployment are mainly dependent on different AI algorithms in order to ensure the capability of automatic learning from the data as well as autonomous \& optimal decision making. However, since many AI algorithms are mainly relied on vehicular data to train the model, the dynamic and heterogeneous natures of IoV scenarios further add difficulties to implement ML techniques. 

Nevertheless, most of the aforesaid aspects are almost common with vehicular ad hoc network scenarios. In such cases, challenges introduced due to these unique aspects and how to meet the requirements, at the same time, how to solve the challenges will be different and unique for IoV application scenarios.

\subsection{Motivations of Using Blockchain in IoV}
Blockchain has the potential to provide a substantial number of innovative solutions to a majority of IoV application scenarios. As such, most of the IoV scenarios are real-time and mobile, and they generate and exchange a large amount of data. Particularly, many classic techniques are unlikely to be suitable and effective in IoV scenarios. Moreover, the increasing connectivity in such scenarios might bring unique attack vectors for the malicious entities. On the other side, the integration of blockchain into IoV not only improves the security, privacy, and trust but also enhances the system performance and automation. Thus, to accommodate flexibility and handle massive data, blockchain-like strong technology should be leveraged. In the following, we present some major stimulating causes for adopting blockchain in IoV.

First of all, decentralization is one of the main features of blockchain. Blockchain allows to create decentralized IoV networks, and it includes more distributed entities, which can be RSUs, vehicles, and people. At the same time, these distributed entities are able to manage their own operations independently. The working principles of current IoV network which is mainly based on central decision will be transferred into decentralized model and become simplified. Ultimately, the decentralization will enhance the user experiences of vehicular services.

Second, blockchain eliminates the dependency of cloud-like systems for data storage and management. Moreover, blockchain along with smart contract enables the removal of third-party entities such as the central service manager, control center, administrators, and trusted intermediaries. Rather, the participants in the blockchain network can maintain the vehicular services and transactions by themselves which will result in reduced operational costs.

Third, security threats such as interruption, single-point-of-failure, and availability attacks can be potentially addressed by adopting blockchain for IoV. This is due to the synchronization and replication of blockchain among all the peer nodes connected to the network. Thus, the services are able to run smoothly even if one or more than one nodes are compromised. On the other hand, the blockchain technology is relied on modern cryptographic techniques to ensure the common security and privacy properties. In fact, blockchain emphasizes the enhanced security and privacy for IoV networks by cryptography.
    
Fourth, blockchain provides high immutability for IoV services and scenarios since in blockchain the blocks maintain a chain to connect with each other through the hash values of each block records. This immutability feature of blockchain potentially prevents data tampering and modification, and also, helps to audit accurately. Additionally, it enables the deployment and enforcement of any pre-defined rules or scripts with the help of smart contract.

Fifth, blockchain offers peer-to-peer (p2p) trading, sharing, and communications between two entities. By p2p network, the service requesters and providers can establish direct communications among them. This p2p feature is particularly useful for IoV scenarios to share data and resources among vehicles and RSUs securely. Since the entities do not need to communication with any intermediary in the p2p network, it ultimately results in low latency applications and services. 
    
Sixth, the IoV connects heterogeneous entities which may not trust each other. Empowered by novel consensus mechanisms, blockchain is able to establish strong trust among even untrusted entities. Besides consensus mechanisms, the smart contract can also play an important role to address the trust issues while making decision without any trusted entity. Moreover, smart contract also helps to achieve an automated and independent system by its scripts.

Lastly, the public blockchain is permissionless and is typically open to all entities. Thus, the use of public blockchain potentially open the door to full access of the data stored in blockchain. It also can enhance the transparency of IoV ecosystem.

\section{Applications of Blockchain-Enabled IoV}
After going through the works presented in the literature on how blockchain technology can be utilized in IoV application scenarios, we determine that blockchain is anticipated to cope with a number of IoV applications and services including vehicular data security, vehicle management, and on-demand transportation services. In the following, we review and investigate on the incorporation of blockchain technology and such IoV application scenarios.

\subsection{Data Protection and Management}
With the gradual increase of IoV applications, more vehicles in IoV network will generate a significant amount of data to enhance the driving safety and improve the vehicular services. In fact, the vehicles will rely on the edge computing nodes like RSUs for their generated data storage \& sharing, management, and utilizing of other shared data. 

However, the edge computing nodes are placed in distributed manner along the roadside which make them vulnerable to security and privacy attacks. Additionally, since these edge nodes are operated by multiple service providers, it is difficult to ensure regulation and trust among them. In fact, the vehicles may not agree to share their data due to the aforesaid concerns.

\begin{figure} [!t]
	\includegraphics[width=\linewidth]{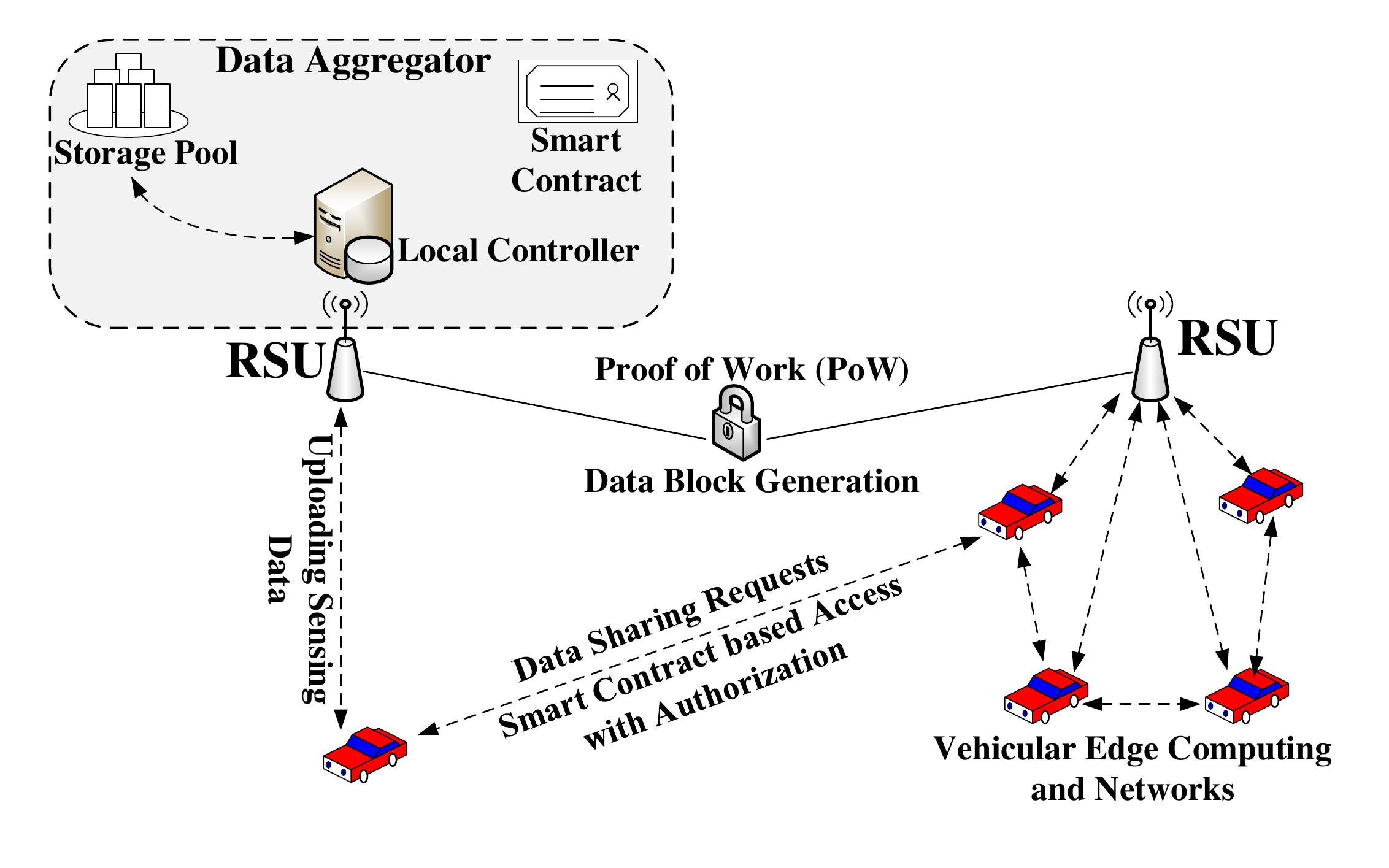}
	\caption{The blockchain-assisted data sharing scheme presented in \cite{kang2018blockchain} for vehicular edge computing networks.}
    \label{fig: vehicluaredgecomputing}
\end{figure}

Recently, blockchain technology has been adopted in vehicular data management scenarios in order to not only address the security \& privacy concerns, but also to establish trust among the edge nodes. For instance, the work presented in \cite{kang2018blockchain} utilizes consortium blockchain, where they introduce a blockchain-based secure and distributed data management system within the vehicular edge computing networks. The advantages from the utilization of smart contract in this proposed system are twofold. First, the smart contract is utilized to achieve secure data sharing and storage within the vehicles as well as the vehicular edge computing servers such as RSU as shown in Fig. \ref{fig: vehicluaredgecomputing}. Secondly, the smart contract also ensures that the data can’t be shared without authorization. Besides, a reputation-based data sharing scheme is developed in this proposed system so that the vehicles are able to choose the optimal and more reliable data source having high-quality data. In order to manage the reputation of vehicles rightly and precisely, a three-weight subjective logic model is employed by taking into account interaction frequency, event timeliness, and trajectory similarity. With this reputation scheme, the proposed system improves the detection rate of malicious and abnormal vehicles over traditional reputation schemes. In the same direction, in \cite{javaid2019drivman}, the authors propose a blockchain-based solution for vehicles, named DrivMan, which facilitates trust management, data provenance, and privacy through smart contract, physically unclonable function (PUF), and public key infrastructure (PKI). By the use of blockchain in the DrivMan solution, the distributed trust management can be maintained in the non-fully trusted network. Additionally, DrivMan achieves data provenance since PUF helps to allocate a unique crypto fingerprint to each vehicles. Furthermore, PKI is employed to enable registration and assign key pairs to the vehicles through a certificate authority (CA). The CA is able to trace the identities of malicious vehicles, and revoke the certificates when necessary. The aim of the PKI is to ensure the privacy and prevent the real identity leakages to the attackers by eliminating the linkability between the identities and respective public keys.

\begin{figure} [h]
	\includegraphics[width=\linewidth]{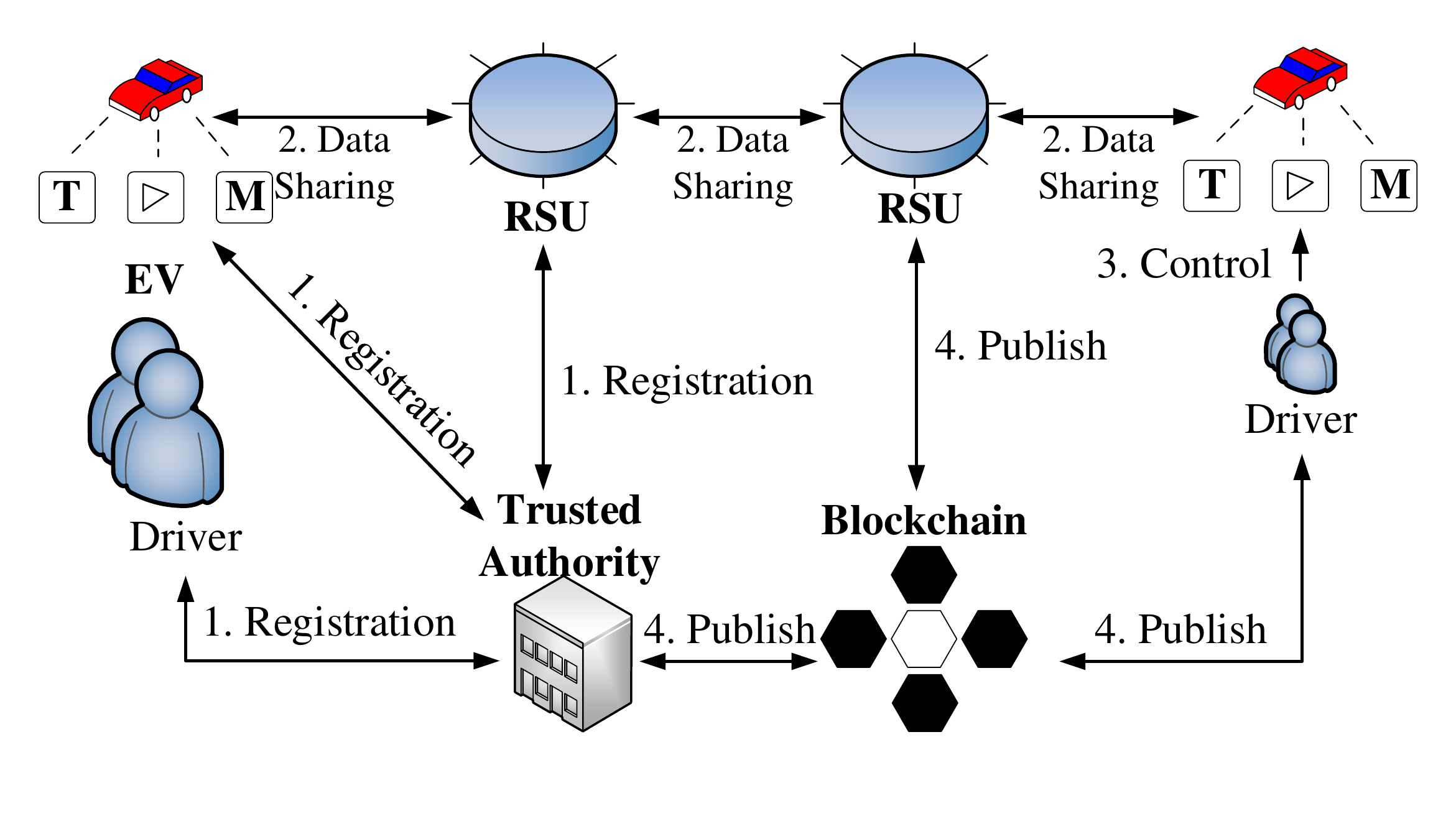}
	\caption{The system model of blockchain-enabled security and privacy-preserving multimedia data sharing introduced in \cite{shi2020blockchain} for vehicular social networks.}
    \label{fig: multimediadatasharing}
\end{figure}

Blockchain is also successfully applied in secure vehicular data sharing in ITS. For example, a multimedia data sharing approach based on blockchain and cryptographic techniques is presented in \cite{shi2020blockchain}, which can be deployed in vehicular social networks. The blockchain is adopted in this approach to leverage its immutability feature to address the challenges of malicious shared multimedia data tampering. Blockchain is also used for data tracing to detect the data sent from malicious users. On the other hand, cryptographic techniques are utilized to protect the privacy leakages of users, vehicles, and RSUs from attackers during the data sharing. The privacy leakages include identities of all entities and sharing habits. Fig. \ref{fig: multimediadatasharing} represents the system model of this approach. Meanwhile, an efficient and security scheme for blockchain-assisted data sharing among vehicles is introduced in \cite{chen2019traceable}. The main contribution of this proposed security scheme is a novel key negotiation scheme which has traceability and authentication properties. Specifically, this scheme aims to effectively address a number of challenges such as shared data security, monitoring, and trust. The proposed scheme offers automated key exchanges which is dependent on static or dynamic scripts in order to enable a fast and automatically executable key negotiations in vehicular communications. It also offers confirmability as well as public traceability of the key negotiation processes with the help of timestamps to safeguard from decryption failure attack. Moreover, the scripts, channel availability, and timeliness of negotiation process ensure defense against the packet dropping attack.

\subsection{Data and Resource Trading}
Data and resource trading in IoV scenarios enables business opportunities for vehicles and other associated entities. But data and resource trading may lead to security issues such as access control enforcement and privacy of users. Furthermore, considering the mobility of the vehicles and wireless link, the trading scenarios could be vulnerable to DoS and jamming-like malicious attacks. In fact, such attacks could be unintentional as well. Another concern is to eliminate any possible disruption due to malicious attacks and also, ensure intermittent as well as trustworthy trading among the trading parties.

Recent works have shown that blockchain technology has emerged as a potential enabler to address most of the aforementioned issues and concerns. Specifically, blockchain could offer secure, peer-to-peer, and decentralized solutions for data and resource trading among multiple entities in IoV context.

\begin{figure} [b]
	\center
	\includegraphics[width=\linewidth]{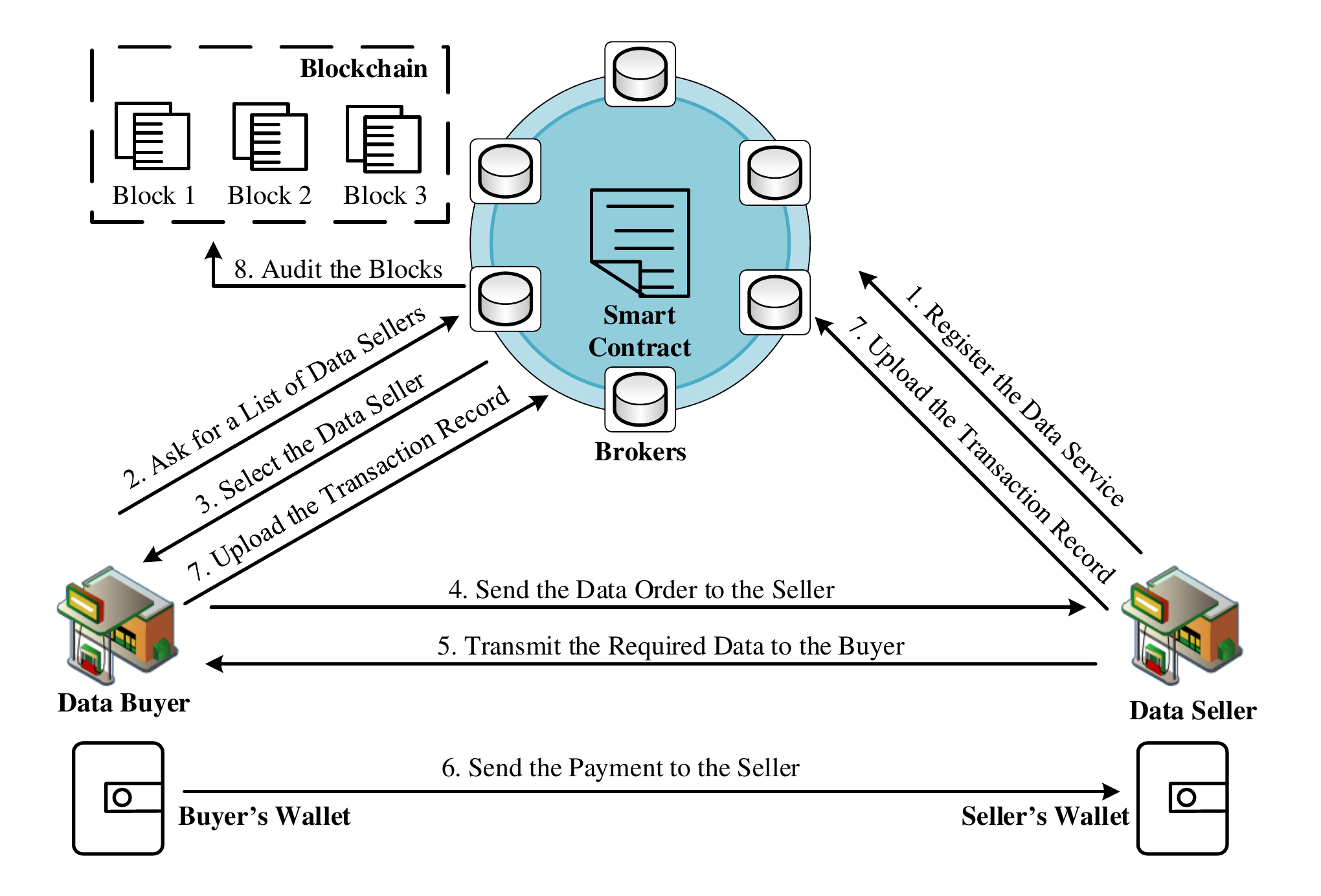}
	\caption{A secure peer to peer data trading architecture utilizing consortium blockchain proposed in \cite{chen2019secure}.}
	\label{fig: peertopeerdatatrading }
\end{figure}

In particular, blockchain technology has been integrated with vehicular data trading to facilitate peer-to-peer trading. As an example, a blockchain-assisted framework for the IoV, which enables secure and efficient data trading is proposed in \cite{chen2019secure}. Here, the blockchain technology is utilized in order to solve the problems of data trading such as lack of transparency \& traceability and unauthorized data modification. The architecture of blockchain in this framework as depicted in Fig. \ref{fig: peertopeerdatatrading } is considered as consortium based, where a number of local aggregators form a consortium to do transactions auditing and verification. In addition to this, an iterative double auction mechanism is adopted in order to optimize the price of data, maximize social welfare while preserving the privacy of both the buyers and sellers so that more users will be encouraged to take part in data trading. Moreover, the data transmission cost is also considered to enhance the system stability.

\begin{figure} [h]
	\includegraphics[width=\linewidth]{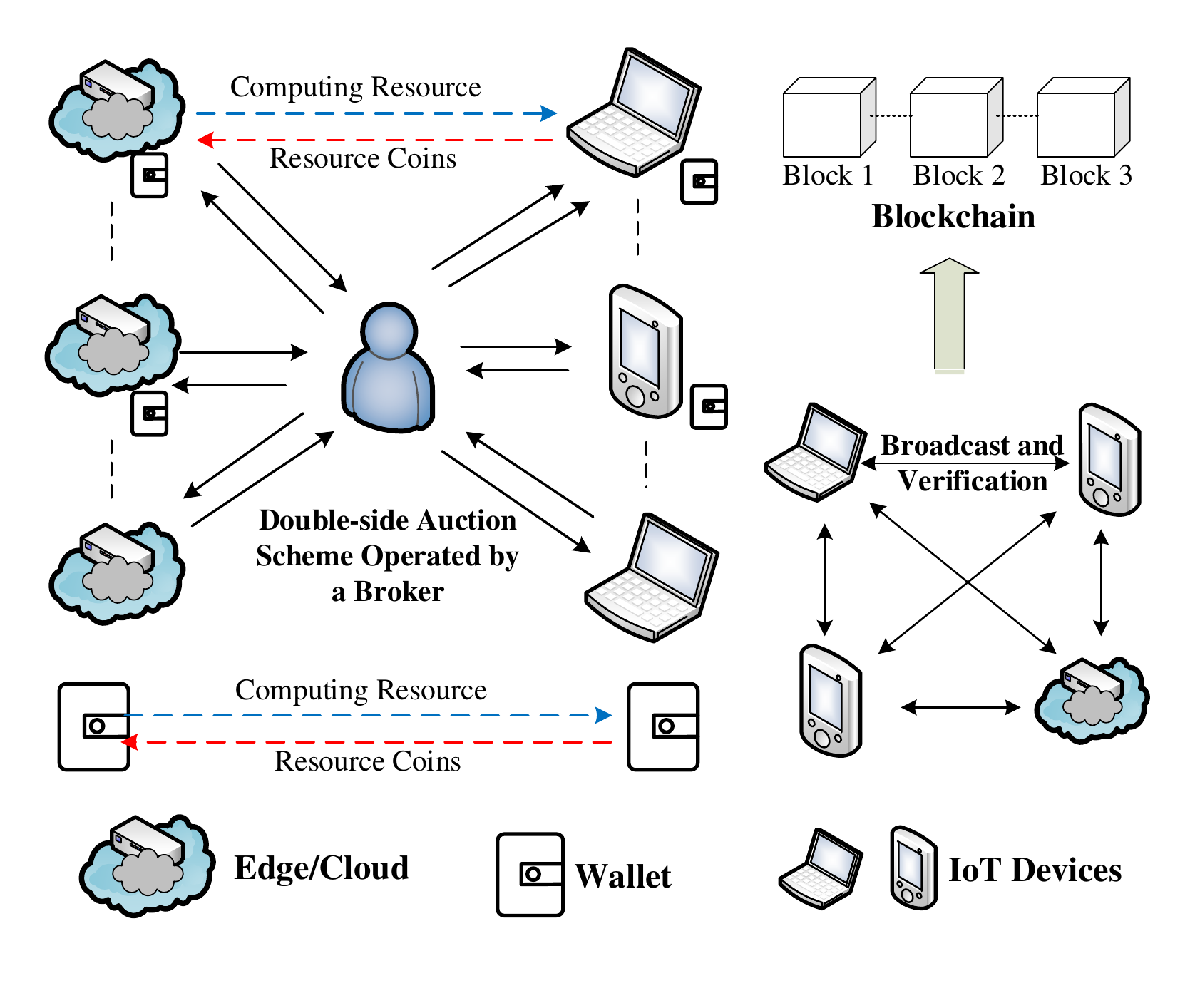}
	\caption{The working flows procedures of proposed blockchain-based computing resource trading platform presented in \cite{li2019computing}.}
	\label{fig: computingresourcetrading}
\end{figure}

\begin{figure} [h]
    \center
	\includegraphics[width=\linewidth]{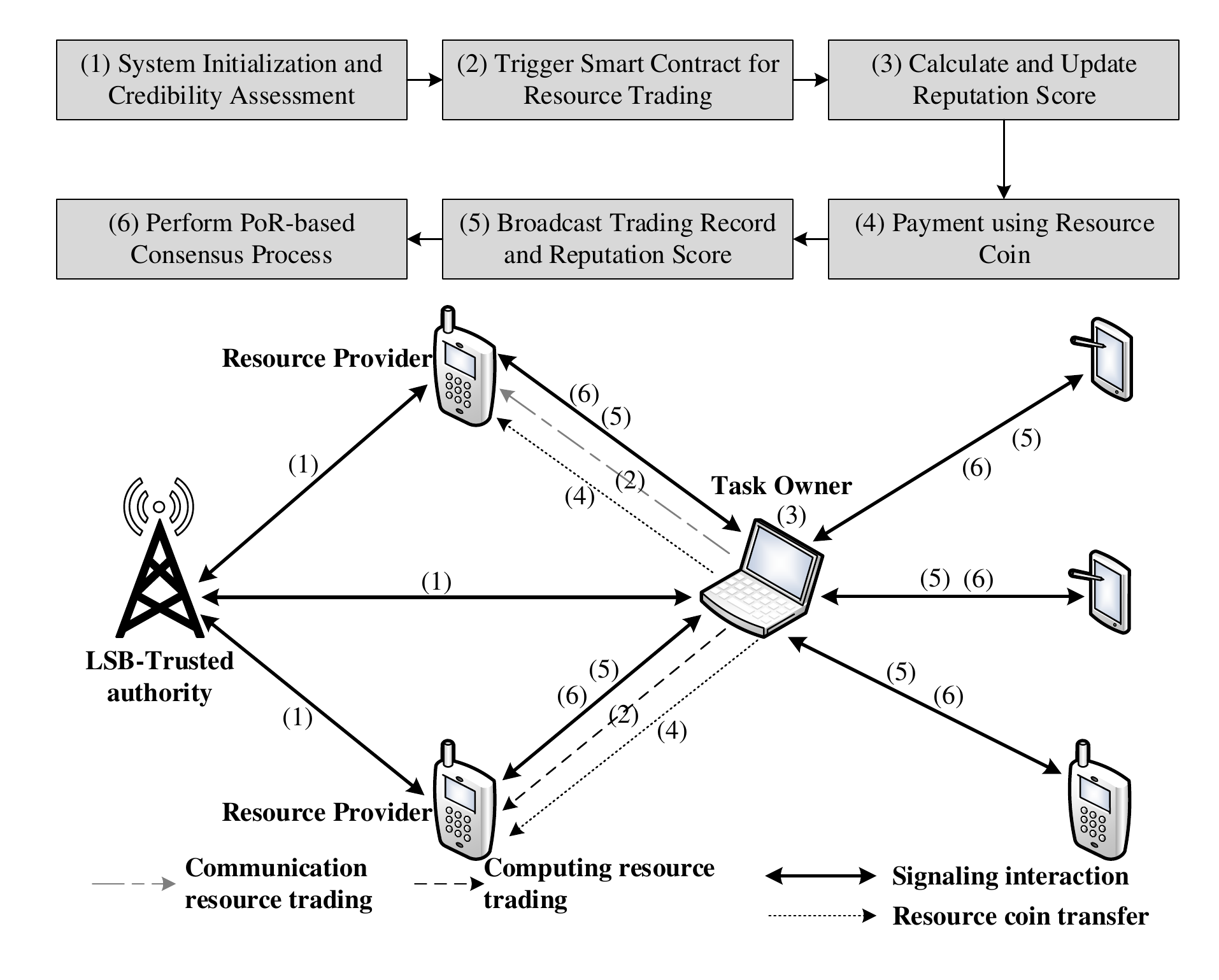}
	\caption{The working procedures of the resource trading solution based on blockchain proposed in \cite{qiao2019blockchain}.}
	\label{fig: resourcetradingsolution}
\end{figure}

In addition, blockchain technology has been incorporated with vehicular resource sharing. For example, a peer-to-peer system using blockchain to provide a computing resource trading platform is presented in \cite{li2019computing}. The aim of this proposal is to address the resource trading problems for edge-cloud-based systems such as ensuring a truthful bid and allowing both seller \& buyer to engage in trading. To maintain the trading market, a broker is presented in this proposal. Next, the allocation problem will be solved by the broker to identify the amount of traded resources and to model a particular price rule to encourage traders so that they offer truthful bids. To do this, this proposal uses an iterative double-sided auction mechanism. As a result, it also ensures maximum social welfare and enables individual rationality as well as balanced budget. Fig. \ref{fig: computingresourcetrading} represents the detailed process of the computing resource trading relied on blockchain. Similarly, a decentralized framework referred as D2D-ECN for IoT applications is presented in \cite{qiao2019blockchain}. This D2D-ECN is a resource trading and task assignment solution which is based on blockchain, smart contract, edge computing, and device-to-device communications as depicted in Fig. \ref{fig: resourcetradingsolution}. The D2D-ECN aims to ensure a collaborative platform where the computational-intensive as well as latency-aware applications can be implemented. Also, it ensures low-latency processing of real-time application scenarios by offloading the computational tasks. To make a balanced latency and decision time, a task assignment approach is developed based on swarm-intelligence. Moreover, it establishes the trustworthiness among the resource service providers and task holders while addressing the efficiency issue of resource management. Additionally, a proof-of-reputation consensus mechanism is introduced instead of PoW for resource-limited devices. In this mechanism, the entity having the higher reputation value can only allowed to pack the resource transactions, and the reputation values are stored in the blockchain. The reputation values for each entities are calculated by considering both the present computational performance and previous history. In this regard, to provide rewards to the participants, a game theory-based mechanism is developed.

\subsection{Resource Sharing}
IoV brings opportunities to develop a cooperative resource sharing platform for both stationary and running vehicles. This platform is able to bring an opportunity to share their spare computational and spectrum resources with their nearby entities. This platform not only assists to implement latency-sensitive services, but also supports AI applications.

However, in spite of such clear benefits, there remains two issues. One issue is that it is difficult to develop and maintain such platform since the vehicles may not trust each other. On the other hand, with blockchain, the trustworthiness can be ensured while vehicles sharing the resources. Another issue is to develop proper mechanisms by integrating with the economic benefits so that the vehicles will be encouraged to participate in resource sharing.

\begin{figure} [h]
	\center
	\includegraphics[width=\linewidth]{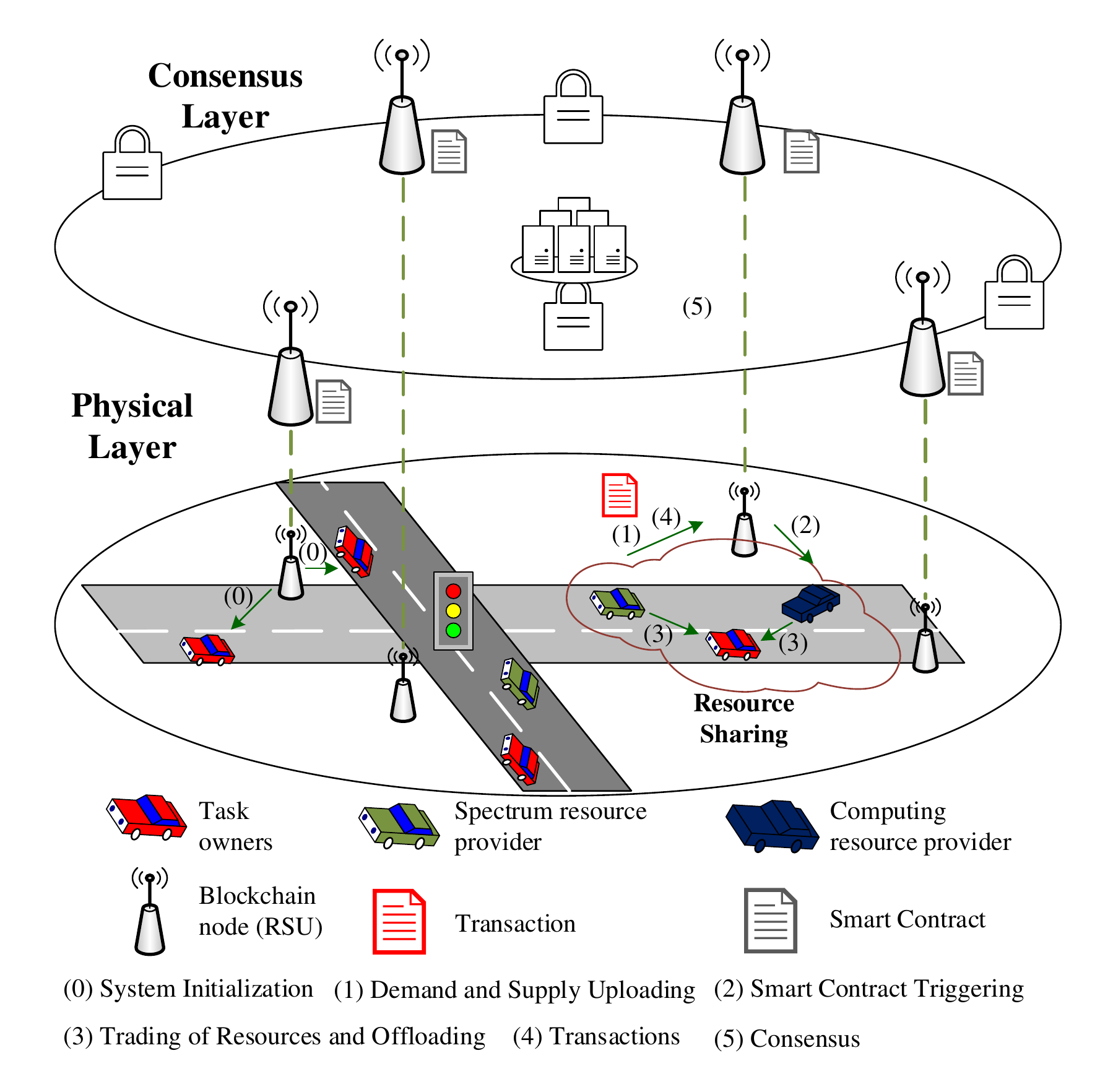}
	\caption{The system components and basic workflow of the resource sharing scheme introduced in \cite{chai2019proof} for IoV.}
	\label{fig: resourcesharing}
\end{figure}

In particular, blockchain is able to support a decentralized platform so that the vehicles can share their resources among others with an aim to increase the efficiency and capability. Indeed, blockchain can address the problems in resource sharing such as establishing trust and at the same time, ensuring security \& privacy of the entities. Such a blockchain-assisted resource sharing scheme for IoV is presented in \cite{chai2019proof}. In this scheme, a reputation-based consensus mechanism which is light-weight compared to most popular PoW is presented in order to establish trust and decrease the dependency of computationally expensive mining process. Here, the reputation scores are employed to show the trustworthy values of vehicles. For trust management, the sharing procedure and the proposed consensus mechanism are incorporated together by using the reputation scores of vehicles. Moreover, this scheme replies on RSUs which are responsible for maintaining the sharing records and blockchain. In addition to these, a smart contract-based resource pricing mechanism is also presented, where a deep reinforcement learning technique is employed, by considering the high mobility characteristics and places of vehicles to facilitate the demand and supply matching while sharing the resources. This proposed pricing scheme for resource sharing is at least 30\% superior to the contemporary unified pricing schemes. Next, the vehicles engaging in resource sharing and block validators are kept separated in this scheme to preserve the privacy of the vehicles and decrease the communication delays of publishing new blocks. Fig. \ref{fig: resourcesharing} depicts how the resources can be shared within the IoV.

Besides, with the help of blockchain, even the non-moving vehicles are able to share their idle computing and networking resources during parking securely and efficiently. The blockchain is adopted in \cite{wang2019permissioned} to develop a permissioned chain for vehicular network named as Parkingchain. An improved DBFT consensus mechanism is developed for Parkingchain which uses a multi-weight subjective logic to calculate the reputation values of the vehicles accurately. The highly reputed vehicles will be assigned as consensus entities to verify and audit the transactions. Moreover, an incentive scheme is also introduced to provide reward to the vehicles with the help of contract theory. However, the secured approach and incentive scheme aim to motivate vehicles ultimately to share their idle resources in vehicular network.

\subsection{Vehicle Management}
Two most popular vehicle management examples are smart parking and vehicle platooning. The smart parking management system allows vehicle users to know about the free parking space information to book in-advance. This pre-booking aims to reduce the wasting times of vehicle users and traffic congestion. On the other hand, the vehicle platoon idea is a technique to combine a number of vehicles together who have almost a common interest. Basically, a vehicle with or without driver leads the other autonomous vehicles, and all vehicles maintain a short space between each other. The leading vehicle is called as platoon head, whereas, the others are considers as platoon members. 

With the access and management of a large number of vehicles, parking lots as well as road-side units, however, the management system will become more complicated. In fact, it will be difficult as well to ensure privacy of the users and large-scale involvements among vehicles, vehicle users, and parking lot owners if the management is still dependent on central system.

The introduction of blockchain into vehicle management systems can essentially address these aforesaid problems. Specifically, blockchain can address the current problems of centralized parking system such as requirement of revealing the private information during booking while searching free parking spaces like destination information and having centralized architecture with possibilities of availability attacks \& data leakages. The authors in \cite{amiri2019privacy} introduce a blockchain-based scheme so that the vehicle drivers are able to search the available parking spaces to book in-advance in a decentralized and privacy-preserving way. In this scheme, the blockchain concept is utilized to eliminate the problems of centralized system. The parking lot owners form a consortium to maintain the blockchain such as sending their parking offers along with information, though they do not trust each other. Next, these parking offers will be kept on the blockchain as record. In addition to blockchain, the private information retrieval (PIR) technique is also used to ensure the location privacy of vehicle drivers. The PIR technique allows drivers to inquiry the parking offers in blockchain network without disclosing their desired destinations. Once retrieved a parking offer from blockchain network, the driver uses a short randomizable signature to confirm the booking directly with the parking owner in an anonymous way. However, this short randomizable signature ensures conditional privacy so that the trusted authority will be able to reveal the real identity of drivers to take necessary actions in case of any malicious activities. Furthermore, an anonymous payment system is proposed so that the drivers can make payment instead of traditional card payment systems which could disclose sensitive information.

\begin{figure} [h]
	\includegraphics[width=\linewidth]{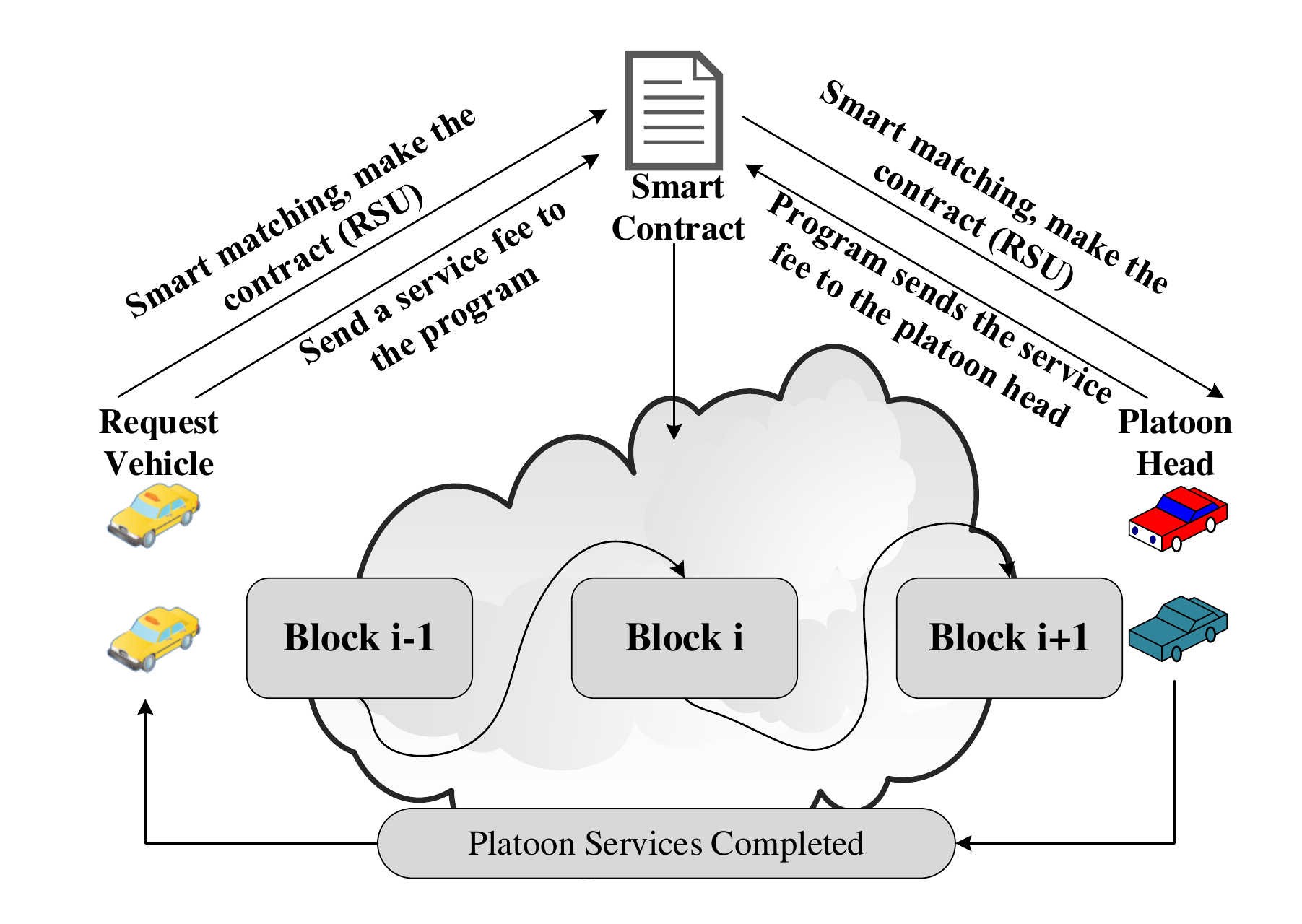}
	 \caption{The components and basic workflows of proposed platooning model introduced in\cite{xiao2019smart}.}
	 \label{fig: platooningmodel}
\end{figure}

In addition, with the openness, decentralization, and security natures, blockchain is able to accelerate the vehicle platooning management. As an example, a model for blockchain-assisted urban IoV is proposed in \cite{xiao2019smart} so that the autonomous vehicles (AVs) can establish a group which is referred as vehicle platoon to expedite the automatic driving and intelligent transportation. The main aim of platooning is for long distance cargo transportation using trucks, and the advantages is fuel saving, due to reduced air friction. There is no payment because the trucks belong to the same delivery company. However, to create a platoon, this proposed model uses path information matching to allow only the selected AVs to have successful matching. The member AVs are driven by a platoon head which is chosen by calculating the reputation value. Besides, a scheme is proposed in this model which is responsible for giving the incentives to the AVs to promote them to become platoon head and also, keep the updated records of platoon members. Application of blockchain and smart contract in this model can solve the problems of false and malicious payments, and offer secure payment among platoon head and member vehicles. The members in a platoon is responsible for paying some fees to the head due to the services in accordance with the contract. The Fig. \ref{fig: platooningmodel} illustrates the basic workflows of this model.

\subsection{Ride Sharing}
The ride sharing services have become popular through some prominent service providers like the Uber due to the convenient usage of travelling. With these services, people can share a vehicle to go to the same destination. A number of people sharing one vehicle ultimately reduces the traffic congestion particularly in busy hours, the carbon emission from vehicle, and the number of vehicles in the roads. Consequently, it also improve the mobility of vehicles in urban area. Moreover, it is expected that future autonomous vehicles will foster the ride sharing services by becoming a part. This kind of sharings are often referred as on-demand transportation services, and with the advancement of smart phones and their applications, such services becoming more available to reach. 

Despite the obvious benefits, the existing cloud-assisted ride sharing services have two unsolved challenges. First, because of the centralized cloud connection, the users may suffer from unnecessary communication delays. Secondly, the centralized storage system has introduced the risk of users’ privacy disclosure and privacy management. Thus, in order to motivate both vehicle users and drivers to share their vehicles, it is necessary to develop an efficient, secure, and privacy-preserving ride sharing service.

Recently, blockchain has been already employed in such context. In \cite{shivers2019toward}, the authors address the difficulties of implementing the ride-sharing platform for autonomous vehicles with the integration of traditional centralized architecture. A blockchain-assisted decentralized ride-sharing framework is developed so that the autonomous vehicles will be able to offer themselves in a community-driven fleet independently. The proposed framework is specifically developed to serve as the intermediary platform between the riders and autonomous vehicles with an increased flexibility concerning the riding price and profit sharing. This framework takes the advantages of blockchain which is maintained by a number of owners of autonomous vehicles who will be working together, where blockchain will establish trust among non-trusting nodes. Most importantly, this proposed framework uses Hyperledger Febric blockchain platform along with chaincode protocol. The purpose of using the chaincode protocol is to ensure data security while client applications connect with the blockchain network. Meanwhile, another new ride-sharing model based on blockchain is presented in \cite{li2019coride}, which is called “CoRide”. The proposed service model offers the following benefits: (i) it introduces a decentralized ride-sharing model based on blockchain, smart contract, PoS, and RSU; (ii) it enables various service providers to provide services collaboratively; (iii) it addresses the limitation of doing privacy-preserving passenger and vehicle matching in centrally by the service provider; (iv) it brings all isolated passenger datasets of different service providers together; (v) it removes the dependency of traditional online trusted party; (vii) it is able to ensure conditional privacy to the passengers, and the identity of a particular passenger is published once all service providers are present; and (viii) It is able to defend the false location attack, malicious RSUs, and possible collision of service providers. Since CoRide brings a number of service providers into one platform, it well fits into the consortium blockchain. In fact, the private proximity test is utilized to authenticate the locations of passengers, and then, set private keys to match and negotiate the passengers. Additionally, a privacy-preserving query processing is also adopted to search the vehicles and match the destinations. Based on smart contract, the proposed model can enforce the RSUs to match the passengers and drivers automatically. Besides these, the recently popular cryptocurrency Zerocash is revised and used to do anonymous payment among the users, drivers, and service providers.

In addition, blockchain is also integrated with carpooling system to offer users in order to share a vehicle efficiently while ensuring privacy. For instance, a blockchain and fog-computing-assisted carpooling system is introduced in \cite{li2018efficient}. This carpooling scheme is referred as FICA (efficient and privacy-preserving carpooling). The FICA has five entities such as a trusted party, users, vehicle drivers, fog nodes, and cloud servers. At the beginning, the trusted party initializes the whole system by generating the public parameters as well as cryptographic keys for the entities. Then, the user who wants to share a vehicle with other users send a carpooling request to nearby fog node. The request consists of mainly a pseudo ID, a key, and an encrypted get-off location. On the other side, the driver receives the request from the fog node which has been sent by the user, and sends a response with free seats information to that fog node. The response of driver consists of a pseudo ID, a number of encrypted get-off locations, and some location proofs. After that, the fog node gathers both request and response from the user and driver respectively. Next, it authenticates the user \& driver, checks the integrity of data, helps the user to match with a nearby driver, and finally, send the carpooling data to the cloud. The responsibilities of cloud server are to gather all requests and responses from different fog nodes, support fog nodes by replying the queries of vehicle traffics, and monitor the conditions of traffics. The fog nodes are utilized in this scheme to address the limitations of cloud computing such as request \& response time delays and extra communication overhead since the users usually search local drivers, though the fog computing nodes are connected with cloud platform. In FICA scheme, the users and drivers are authenticated by using anonymous authentication technique to ensure privacy from both cloud and fog nodes. The anonymous authentication also offers conditional privacy so that the identities of malicious users can be recoverable by trusted authority. Moreover, FICA scheme also offer one-to-many matching and destination matching by using a private proximity test and a range query techniques respectively. in particular, a private blockchain is developed based on PoS for this proposed scheme which is maintained by the fog nodes. The aim of this blockchain is to store carpooling records in a verifiable ledger to ensure data auditability. This private blockchain along with trusted party enables carpooling records reliably as well as refrain from any malicious activities by users.

\subsection{Contents Broadcasting}
Contents broadcasting among the Internet connected vehicles aim to improve the in-vehicle services as well as safety services. These services are able to offer a number of attractive applications such as advertisement publication, online shopping, promoting commercial products, entertainment services, and emergency information announcements. Additionally, the edge computing nodes like RSUs can play an important role to support the vehicle services by caching popular contents, broadcasting the contents to nearby vehicles, and reducing the latency in order to enhance the experiences of vehicle users. 

However, blockchain has great potential to enable p2p content sharing and broadcasting among the vehicles through V2V and V2I communications. Indeed, blockchain-assisted p2p model can ensure an efficient, in-expensive, and trustworthy platform for the IoV. Furthermore, to accelerate the broadcasting of contents, blockchain enables to develop novel reward schemes.

Recently, blockchain has been incorporated in vehicular content broadcasting scenarios. For instance, the work in \cite{li2019toward} focuses on commercial advertisement publishing, to allow the users to promote their products in vehicular networks through V2V or V2I. A blockchain-based decentralized advertisement publishing scheme is proposed to realize security, privacy, and fairness. This scheme also utilizes the smart contract, Merkel tree, and zero-knowledge proof techniques. The smart contract and Merkel tree are responsible for maintaining the fairness property between the advertiser and vehicle by achieving the “proof-of-ad-receiving”. The proof-of-ad-receiving concept is introduced to confirm whether a vehicle obtains an advertisement without any cheating. This concept is introduced to mitigate the possibility of obtaining rewards without publishing any advertisement by deceiving the advertiser. Also, this concept allows to detect and punish any advertisement publisher vehicle who maliciously claims reward more than once. On the other hand, the zero-knowledge proof technique is utilized to protect the privacy of vehicles in terms of anonymity and conditional linkability while engaging in the process of advertisement publishing. In particular, this proposed blockchain-assisted decentralized scheme increase the service availability by protecting against external DDoS attacks as well as internal single point of failure/compromise. In the same direction, an event-driven message (EDM) protocol via blockchain, edge computing, and 5G is proposed in \cite{nkenyereye2020secure} for vehicular application scenarios. The EDMs are different type of traditional messages which are used for emergency warning announcements only, and these messages are usually produced during any accident or emergency time. The application of blockchain in this proposal serves to ensure distributed storage and auditability of EDMs. In order to ensure reduced response time, edge computing is adopted in this proposal to process EDM locally. Moreover, 5G communication is considered to provide scalability and low latency services to the vehicles. In fact, this proposal solves the high processing time, security, and access control challanges by utilizing a lightweight, efficient, and pairing-less multi-receiver signcryption technique.

\subsection{Traffic Control and Management}
Based on vehicle generated information such as traffic data, accident incidents, and sharing the information with others will make traffic management related services easier than before. The utilization of the information can bring a number of opportunities and advancements for vehicles in smart transportation system such as dynamic traffic control \& management, traffic conditions monitoring, and traffic congestion mitigation.

However, such use cases expect secure control and management. Particularly, the security is a significant issue. In fact, the transfer and storage of vehicular traffic related data from the moving vehicles to the stationary road side units might be vulnerable to security attacks including breaking the data availability, integrity, and privacy. Another challenging issue is how to develop a autonomous traffic control and management system.

To cope with the aforesaid issues, blockchain can be a better solution which can support four aspects, namely, decentralized management, availability, automaticity, and immutability. As such, a semi-centralized traffic signal controlling approach is introduced in \cite{cheng2019sctsc} which is referred as SCTSC (Semi-Centralized Traffic Signal Control) system. The aim of this system is to support automatic traffic signaling which is an emerging IoV application to enhance the efficiency of traffic controlling. This proposed system collects data from vehicle sensors and drivers. Compared to currently popular pretimed mode, the proposed approach aims to offer control-by-vehicles approach to manage and control the traffic lights dynamically though the collected data. The attribute based encryption (ABE) technique and blockchain are utilized in this system. The ABE technique ensures access control in the traffic related data where the vehicles are divided into groups according to their attributes like locations and destinations so that this system works better than pretimed approach in terms of efficiency and vehicle passing rate. On the other hand, blockchain enables the vehicles to reach a common agreement through a voting process where they decide the signal timing. Next, the agreement and decision will be stored on blockchain. Here, the vehicles will be anonymous while communicating with each other, in order to keep their privacy. Meanwhile, to realize an intelligent traffic system, the work presented in \cite{ren2019intelligent} proposes a system for smart transportation. This proposed system employs blockchain with IoT connected sensor nodes deployed in vehicles. A credit-token method is also presented so that the vehicles will be able to pay the service usages and misuses. Compared to other systems, blockchain is utilized in this work to decrease the traffic congestion and enhance the road safety.

\subsection{Forensics Application}
Assisting in digital forensics is one of potential blockchain applications. The forensic application in IoV includes traffic accident analysis particularly for autonomous vehicles. 

\begin{figure} [b]
	\center
	\includegraphics[width=\linewidth]{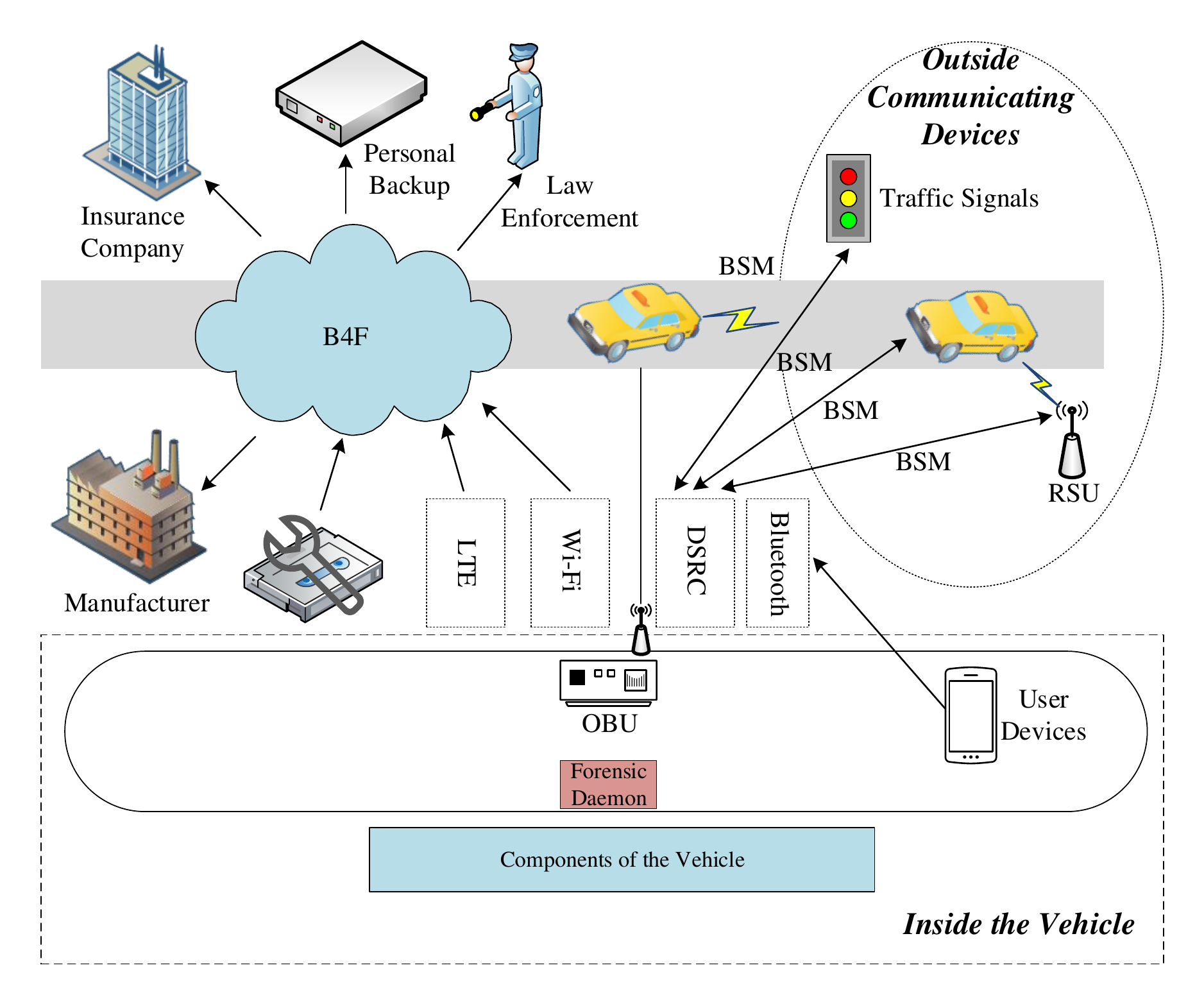}
	 \caption{The forensic model presented in \cite{cebe2018block4forensic} for connected vehicles based on blockchain.}
	 \label{fig: forensicmodel}
\end{figure}

In this context, blockchain can offer autonomy while calculating the post-accident demurrages through the smart contract, bring all parties together, and keep all vehicle related data as a records. A comprehensive vehicular forensic framework named Block4Forensic using permissioned blockchain and vehicular public key infrastructure (VPKI) is presented in \cite{cebe2018block4forensic}. The purpose of this Block4Forensic framework is to provide forensic service to investigate vehicle accidents. In this framework, the blockchain brings vehicles, road-side units, vehicle manufacturers, insurance parties, and police agencies into a common platform, as illustrated in Fig. \ref{fig: forensicmodel} where the evidences and other related data are collected and kept in a blockchain. Fragmented distributed ledger and shared ledger techniques are proposed to store the hashed data and detailed vehicle related data, respectively. The reason behind the shared ledger is to reduce the storage and processing overheads. On the other hand, to ensure the membership establishment, VPKI is adopted. Moreover, this framework allows the vehicles to protect their identity from unauthorized users by using pseudonyms. Another work presented in \cite{pourvahab2019digital} proposes a new blockchain, smart contract, and software defined networking (SDN) based architecture to achieve digital forensics and data provenance. This proposed architecture aims to address the limitations of centralized forensic management, such as integrity and reliability. To enhance the integrity and reliability of forensic data management, blockchain is utilized to store and collect the digital evidences. This architecture also considers cloud computing, where once an evidence is received by the cloud, the SDN controller records it as a block. Moreover, this architecture is baed on modern cryptographic techniques. For instances, it not only utilizes secure ring verification assisted authentication for ensuring authorized user access, but also sensitive-aware deep ECC for data confidentiality. Apart from these, a fuzzy smart contract is implemented in the blockchain to maintain data provenance by tracing the data.

\begin{table*}[h!]
    \centering
    \caption{Summary of blockchain application scenarios in the IoV}
    \label{tab: bcapplication}
    \begin{tabular}{m{1.7cm}|m{5.5cm}|m{8cm}|m{1.2cm}}
\hline \hline
Application Scenarios & Short Descriptions & How Blockchain Brings Opportunities & Papers \\
\hline
Data Protection and Management & - The vehicles share their generated data to their nearby peers, and also, utilize the stored and shared data

- The dependencies of distributed RSUs and vehicles for vehicular related services result in the difficulties to ensure security and privacy such as data integrity, data availability, and privacy of the vehicles while ensuring the trust among the RSUs without relying any third party & 
- The works of blockchain-assisted IoV have showed that blockchain offers decentralization of vehicle data management which eliminates centralized management and control approaches through replication and consensus mechanisms 

- Blockchain has been utilized in vehicle data protection due to its immutability and decentralization features to tackle the data integrity and availability problems

- It has offered much more efficient vehicular data management compared to traditional database management system while ensuring security requirements through modern cryptographic techniques & \cite{kang2018blockchain, chen2019traceable, shi2020blockchain, javaid2019drivman} \\
\hline
Data and Resource Trading & 
- The IoV allows vehicles to trade their generated data as well as own computing and communication resources through Internet, where the vehicles desire direct and independent trading among others &	
- Blockchain has been utilized to establish peer-to-peer (p2p) data as well as resource trading among the vehicles where trading and payment facilities are deployed in decentralized ledger network which is maintained by only the users like vehicles

- This independent trading has helped to decrease the transaction costs significantly & \cite{chen2019secure, li2019computing, qiao2019blockchain} \\
\hline
Resource Sharing & 
- The cooperative resource sharing among the vehicles potentially improve the latency-aware and AI related services 
- It is challenging to ensure direct resource sharing among the vehicles with security and proper incentives due to the trust and mobility issues & 
- To expedite the resource sharing among the vehicles, blockchain has eliminated the reliance of trusted third parties, brokers, and intermediaries

- Blockchain offers direct resource sharing between service provider and user which ultimately results lower latency services and increase the overall performance of services & \cite{wang2019permissioned, chai2019proof} \\
\hline
Vehicle Management & - Due to the heterogeneity and a large number of entities involved in the network to provide smart parking like services, it is challenging to preserve privacy of users as well as implement the access control while the system is under centralized controlled

- The IoV network becomes more complex to manage & - The use of blockchain has been potentially enhanced the vehicle managements in smart parking and vehicle platooning systems by its decentralization feature

- Vehicle management over the decentralized network has fostered the verification of the service requests and contributions through the smart contract & \cite{amiri2019privacy, xiao2019smart} \\
\hline
Ride Sharing & 
- It is becoming a trend to integrate the vehicles and users into ride sharing services through the Internet which are typically based on cloud platform 

- Outsourcing the information of users to the cloud can compromise the vulnerable privacy of users & 
- Blockchain has offered better ride sharing services compared to cloud-based services in terms of lower communication delay and quicker data retrieval

- It has offered vehicles and passengers searching and matching locally through RSUs instead of passing cloud platforms, which simplify the ride-sharing services & \cite{li2018efficient, shivers2019toward, li2019coride} \\
\hline
Content Broadcasting & 

- Content broadcasting is the capability of IoV system to disseminate both non-safety and safety contents through vehicles and RSUs which may be semi-trusted, non-trusted, and attack-prone 

- The reliance of the vehicles and RSUs is put risk of being modified the contents maliciously & - Blockchain has been adopted to ensure trustworthiness of the contents

- Keeping blockchain that contains the content records in multiple nodes has facilitated fast publishing of contents

- The immutability feature has made the contents strong resistance against data integrity attack & \cite{li2019toward, nkenyereye2020secure} \\
\hline
Traffic Control and Management & 
- The vehicle traffic management system is dependent on data generated by vehicles as well as RSUs for storage

- This system may be vulnerable to security and privacy & 

- Different from traditional one which often utilizes centralized control for traffic management, blockchain utilization in traffic system has showed that blockchain has provided a decentralized management and control with consensus mechanisms

- The strong immutability and multiple storage features of blockchain have enhanced the accessibility but still ensured better security protection against denial of service (DoS) attacks

- Smart contract implemented in blockchain has been beneficial to traffic management and control in terms of authentication and authorization without relying on trusted key generation centers and autonomous detection of malicious vehicle users & \cite{cheng2019sctsc, ren2019intelligent} \\
\hline
Forensic Application & 
- The vehicles, in particular autonomous vehicles desire automated while ensuring data integrity for post-accident forensics & 

- Blockchain adoption in forensics application has showed a promising alternative to store the vehicle related records permanently with an aim to prevent malicious tampering and accurate auditing

- The research works have demonstrated that utilizing blockchain along with modern cryptographic techniques forensic application systems enhances traceability and transparency

- The smart contract implemented in blockchain facilitates a decentralized fair agreement among the vehicles, insurance companies, and other related parties, which provides proper deployment of rules & \cite{cebe2018block4forensic, pourvahab2019digital} \\
\hline
\end{tabular}
\end{table*}

\subsection{Remarks}

\begin{figure*} [h]
	\includegraphics[width=\linewidth]{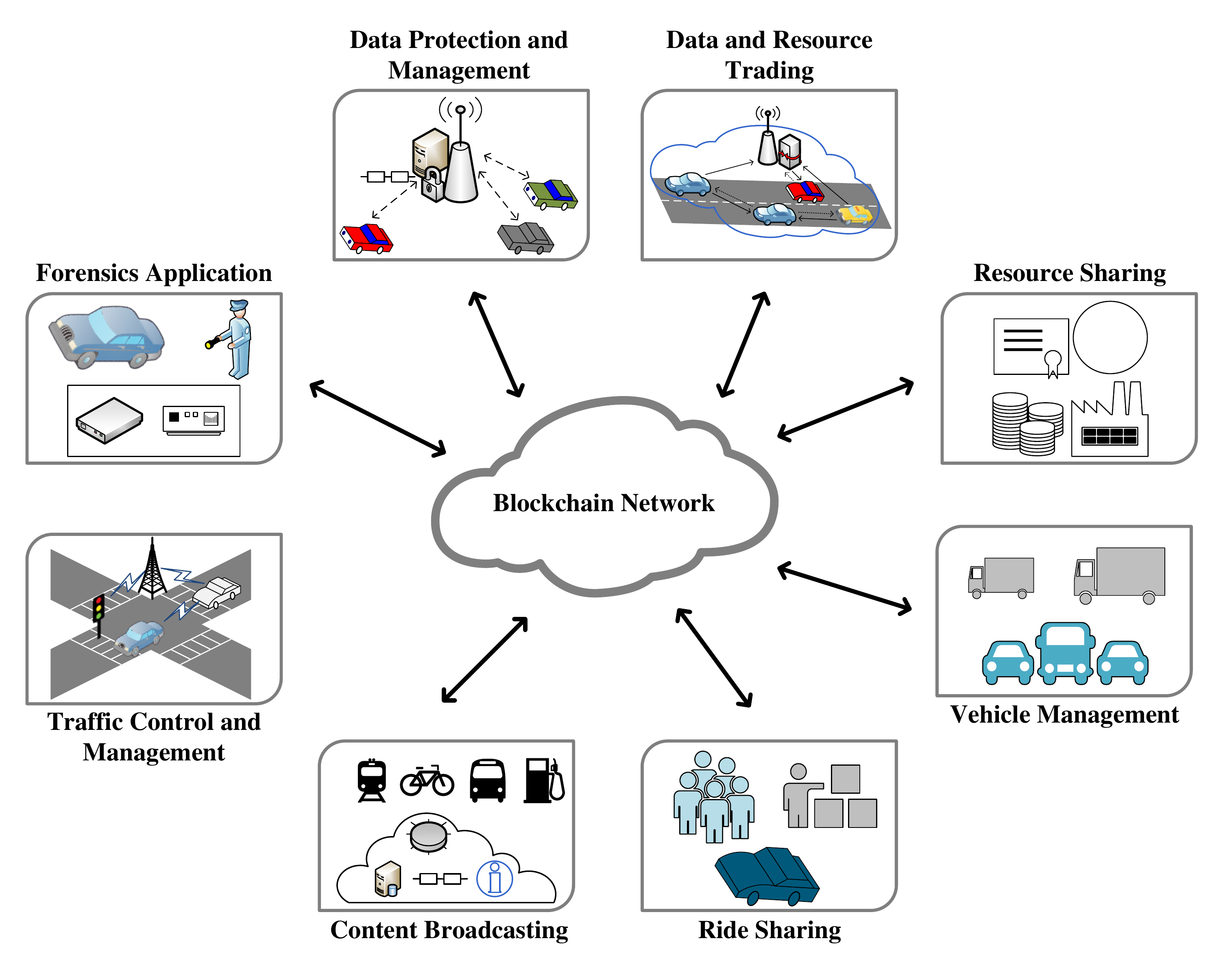}
	 \caption{Integration of blockchain and IoV.}
	 \label{fig: iovapplications}
\end{figure*}

In the aforesaid studies, we have presented a number of blockchain-assisted IoV application scenarios which we have divided into eight major categories such as (i) data protection and management, (ii) data and resource trading, (iii) resource sharing, (iv) vehicle management, (v) ride sharing, (vi) content broadcasting, (vii) traffic control and management, and (viii) forensics application. In Fig. \ref{fig: iovapplications}, these different applications are highlighted. Although the emergence of IoV offers a wide range of applications, we have discussed only the blockchain efforts in IoV. These aforesaid IoV application scenarios are presented in Table \ref{tab: bcapplication} as a summary in terms of descriptions, how blockchain brings opportunities to the applications, as well as related papers.

In fact, blockchain for IoV is attracting more and more attentions in academic research, at the same time, it is also attracting from industries and is expected that it will continue. For instances, according the IBM's report \cite{misc4}, blockchain would be utilized for mobility services such as managing vehicular data and making personalized preferences. Also, according to report by the Deloitte in \cite{misc5}, blockchain is one of the accelerating technologies in automotive industries and it has opportunities across the industry covering from manufacturers to end-consumers.

However, we have assumed that the papers referenced in TABLE \ref{tab: bcapplication} only cover the potential IoV applications with blockchain. Different from this section, in the following section, we will present some architectures and frameworks where blockchain and IoV are incorporated successfully.

\section{Blockchain for IoV Architectures and Frameworks}
Since blockchain offers a number of benefits to IoV applications, recently some blockchain-assisted architectures and frameworks have been introduced for IoV ecosystem. In this sections, we present the most representative architectures and frameworks that integrate the blockchain technology with IoV applications.

\subsection{Blockchain-Based IoV Architecture}

The authors in \cite{jiang2018blockchain} present a blockchain-integrated architecture and network model for Internet of Vehicles (IoV) applications. The main contribution of this proposed architecture is to offer distributed big data storage and security management by blockchain. In this architecture, the data is classified into different types according to the sources and application functions. It includes monitored driving data, vehicle sensor data, passenger data, vehicle insurance data, and transactions.

\begin{figure} [h]
	\center
	\includegraphics[width=\linewidth]{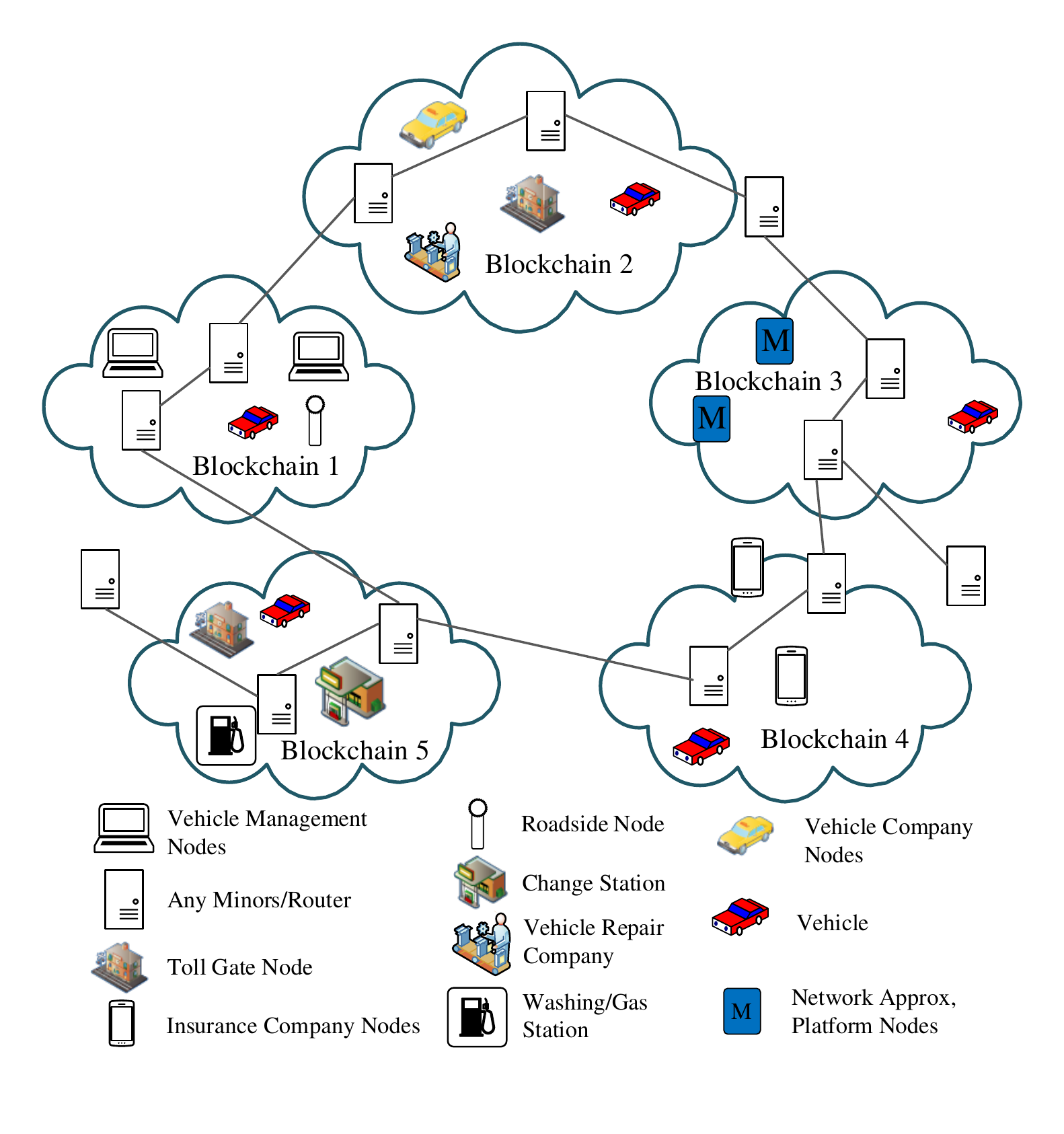}
	 \caption{Network architecture of the blockchain-based IoV presented in \cite{jiang2018blockchain}.}
	 \label{fig: blockchainiov}
\end{figure}

A wide range of vehicles, roadside edge computing units, toll stations, fuel stations, and insurance authorities are connected with the blockchain networks which are responsible for collecting the data and maintain blockchain networks. Based on these variety of data, the blockchain is also divided into five different types of chains as presented in Fig. \ref{fig: blockchainiov}. Here, the five blockchains are considered as sub-blockchain networks, and they are independent to each other with different application purposes. The responsibilities and generation sources of these sub-blockchains (SB) are : (i) SB1: Road-side nodes create blocks which contain data for this SB1, and SB1 is responsible for sharing the blockchain with the neighbor nodes; (ii) SB2: Vehicles generate the new blocks, and SB2 shares the blockchain with road-side units; (iii) SB3: Road-side nodes also create data block for this SB3, and SB3 is transmitted among the neighboring road-side nodes; (iv) SB4: Once road-side units collected data, they will be shared among the nodes situated a toll stations; and (v) SB5: Different service provider stations such as fuel filling, washing, and charging nodes create the blocks and also, maintain the SB5 by themselves. Since the vehicles are not stationary, according to the proposal, the vehicles are either dependent on the surrounding nodes to transmit their generated new blocks, or use the wireless network when they cannot find any nearby neighbor node.

\subsection{Blockchain-Assisted EVs Cloud and Edge Computing}
\begin{figure} [b]
	\center
	\includegraphics[width=\linewidth]{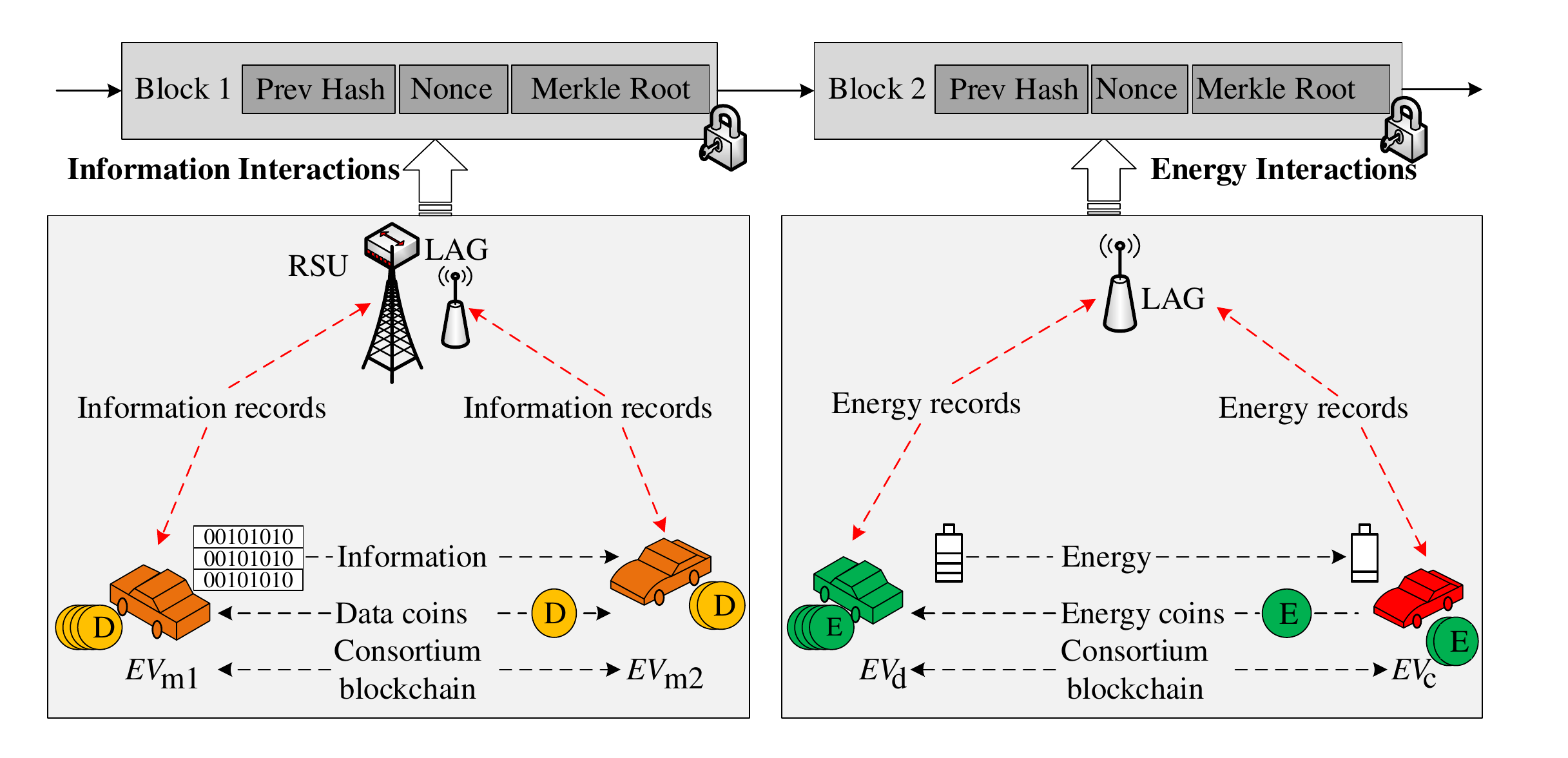}
	 \caption{The structures of two proposed cryptocurrencies in EV cloud and edge architecture based on consortium blockchain presented in \cite{liu2018blockchain} for the purpose of asset balancing \& demand response tracking.}
	 \label{fig: cloudandedge}
\end{figure}

The authors in \cite{liu2018blockchain} present a blockchain-based security solution architecture for Electric Vehicles (EVs) hybrid Cloud and Edge Computing (EVCE). This proposed security solution enables context-aware applications for EVs through energy and information exchanges. Based on blockchain, two different coins are introduced named energy coin and data coin for energy and information interactions respectively. The blockchain structure of these two coins are represented in Fig. \ref{fig: cloudandedge}. The coins aim to provide V2E services such as V2G, V2I, and V2V in order to gain information intelligence and energy sharing while ensuring improved security protection. The purpose of using blockchain is to accommodate two of its two features such as decentralization and distributed consensus into EVCE.

The architecture is hybrid which is different from traditional one, and it consists of two computing modes such as EV edge computing and EV cloud computing. The EV edge computing mode aims to assist the distributed EVs by analyzing and processing a large amount of data while doing collaborative operations. This mode utilizes a number of vehicular cloudlets which are distributed in the network, and focus on the processing of information and energy locally by collaborating with neighbor nodes. The vehicular cloudlets are mobile cloudlets formed with the help of vehicular ad hoc networks which are able to increase the connectivity of network and provide high quality of services. On the other hand, the cloud computing mode is responsible for aggregating all vehicular energy and computational resources in a common pool which are idle to offer cooperative services. Unlike traditional cloud platform, this cloud interact with edge nodes such as road-side units and local aggregators to provide services to the connected and mobile EVs.

\subsection{Blockchain-Centric Automotive System Architecture}
\begin{figure} [h]
	\includegraphics[width=\linewidth]{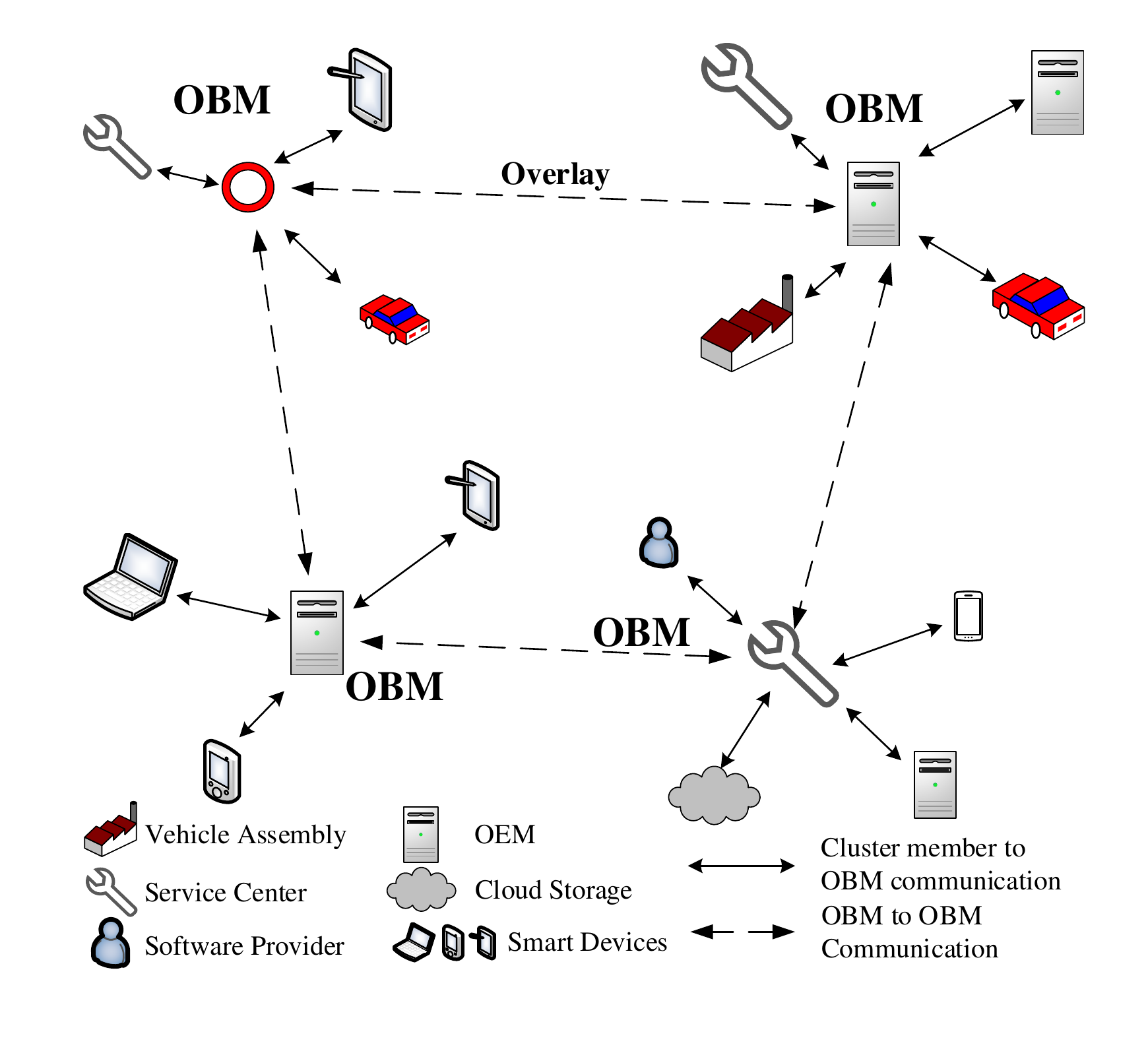}
	 \caption{The overlay network of the blockchain-centric architecture for automotive security and privacy introduced in \cite{dorri2017blockchain}.}
	 \label{fig: automotivesecurity}
\end{figure}

The work in \cite{dorri2017blockchain} proposes a blockchain-centric architecture that enables security and privacy of smart vehicle ecosystem. As shown in Fig. \ref{fig: automotivesecurity}, this architecture consists of three types of entities such as smart vehicles, equipment manufacturers, and service providers. These entities jointly develop an overlay network to communication with each other. Moreover, these entities in the overlay network are also divided into clusters, and each cluster contains cluster head which has responsibilities to manage the blockchain network under it, broadcast the transactions, and verify new blocks. Hence, the cluster heads are considered as overlay block managers (OBMs). In this way, the blockchain technology removes the dependency of centralized control in this architecture. To ensure privacy, each vehicle has an in-vehicle storage in order to keep the privacy-sensitive data such as location traces. The architecture has access control option which allows the vehicle authority to define that which data will be shared with others and which information will be kept in its own in-vehicle storage. The other security and privacy properties for this architecture are mainly inherited from blockchain technology.

\subsection{Privacy-Preserving and Decentralized IoV Architecture}
The authors in \cite{ma2019blockchain} propose a secure, privacy-preserving, and decentralized IoV architecture using blockchain and delegated PoS. This proposed architecture consists of vehicles, sensors and actuators, RSUs, and cloud computing nodes. The vehicles basically generate the data and communicate with other vehicles and RSUs, and are relied on blockchain network to utilize the blockchain data. The vehicles also provide feedbacks according to the service quality as a ratings of the other vehicles and RSUs. The ratings will be broadcasted to other nodes and considered as trust levels. The sensors and actuators help to generate data and control the moving parts of vehicles respectively. Here, the RSUs are considered as major blockchain storage nodes in this architecture to maintain the blockchain network by storing the latest blockchain as well as confirming new blocks. The cloud computing nodes are responsible for storing the blockchain and other data as a backup. 

A hierarchical data exchange approach is also presented in the architecture where it contains two different sub-chains named InterChain and IntraChain to provide access control flexibility to the users. The InterChain enables to share information among vehicles, road-side units, and other infrastructures through V2R and V2V communications. On the other hand, IntraChain maintains the communications within sensors, actuators, and people in the vehicles. 

This proposal makes the blockchain more lightweight and efficient for resource-constrained entities through a reconfigured and optimized structure which can adapt with the vehicular network. The delegated PoS is adopted in InterChain so that the participants are able to reach an agreement on their exchanged and aggregated data. Additionally, to avoid internal collusion of participants, the delegated PoS employs multi-weight reputation technique.

\subsection{Remarks}
In this section, we have presented four blockchain and IoV integrated architectures and frameworks. However, such contributions will not be made into industrial (product) level as well as practical usages as long as the throughput (transactions per seconds) of the introduced works will not be enough to support the IoV applications and services. Besides, blockchain technology’s openness and transparency introduce many participants for contribution. Thus, novel business models are necessary for blockchain assisted IoV ecosystem so that all participants will be treated fairly according their contributions. In this way, the novel business models will be able to support on obtaining the full advantages of the resources of contributors.

Apart from this, since blockchain was introduced in monetary applications initially, adopting blockchain into non-monetary domain like IoT introduce various challenges due to having different characteristics. Furthermore, IoV components have some unique characteristics not present in other IoT applications as discussed in section III-A. Thus, in spite of the functional features that blockchain offers to the IoV applications, it faces some challenges upon deployment which need to address before implementation. As such, we proceed to investigate the challenges in the following section.

\section{Blockchain \& IoV Integration Challenges}
In this section, we discuss about the main challenges that need to be addressed when incorporating blockchain into the IoV scenarios, according to our observation and study. As such, in each subsection, we will first present a short introduction of these challenges, and then, highlight the related works which cover the challenges. Note that in this section, we do discuss about blockchain and smart contract challenges. Based on the existing literature, we have identified four challenges. In the following, we proceed to discuss these four challenges.

\subsection{Security and Privacy}
Integrating blockchain into the IoV offers security and prevention of data manipulations by its ability to guarantee the data immutability. However, blockchain cannot directly guarantee the security and privacy, because blockchain is based on different techniques. Modern cryptographic techniques, pseudonyms, and off-chain storages are some of the techniques which can ensure the security of the contents (transactions or records) inside the blocks as well as the privacy of the users and devices.

To address the aforesaid challenges, a number of works have been introduced to ensure the security and privacy in blockchain-enabled IoV. For instance, to deal with defending the spreading of forged messages as well as ensuring the privacy of vehicles, a work is presented in \cite{lu2018privacy}. To satisfy these goals, an anonymous reputation system is designed which helps to establish a privacy-preserving trust management in vehicular networks. However, to defend the spreading of forged messages, this system takes advantages of a reputation evaluation algorithm, which is able to provide the trustworthiness of messages with the help of assigned reputation values of the vehicles. These reputation values are calculated by considering the historical interactions of vehicles as well as the indirect votes from other vehicles, and these reputation values will be stored in blockchain. On the other hand, to ensure privacy and protect against tracking attacks, instead of the real identities of vehicles, pseudonyms are used in V2I and V2V communications. The public keys of individual vehicles are considered as their pseudonyms. However, the pseudonyms provide conditional privacy so that in case of any disputes, the real identities will be revealed to the trusted authority to identify the vehicles. In another related work, a blockchain-assisted security framework for heterogeneous network of intelligent transportation system is proposed in \cite{lei2017blockchain}. Here, the blockchain technology is adopted in order to expedite the distributed key management in heterogeneous network to obtain better efficiency. This framework consists of two schemes, namely, blockchain-assisted novel key management scheme and dynamic transaction collection scheme. In key management scheme, after eliminating the central manager, a number of security managers (SMs) are introduced to play an important role, i.e,  to verify and authenticate the key transfer processes. The processed records will then be stored in blockchain and shared among the SMs. On the other hand, a dynamic transaction collection scheme enables the system to reduce the key transfer time of blockchain network during the vehicles handover, and the collection period can dynamically change in accordance with various traffic levels. The other responsibilities of the SMs are to capture the vehicle departure information and execute rekeying to the vehicle in the same network.

The limitation of presently available authentication mechanisms which are mainly developed for cloud-based vehicular networks is discussed in \cite{kaur2019blockchain}. Basically, in most cases, these authentication mechanisms fail to guarantee secure \& reliable communications as well as optimal QoS to the highly mobile, latency-sensitive, and decentralized scenarios in vehicular networks. To address this limitation, the authors introduces a decentralized authentication and key exchange mechanism for vehicular fog computing scenarios. This mechanism is based on blockchain along with elliptic curve cryptography (ECC). Here, the blockchain is utilized as usual to maintain the information of networks, whereas the ECC is adopted to ensure mutual authentication between the vehicles \& fog nodes, user anonymity, and reauthentication to engaging vehicles. Both blockchain and ECC will enable the vehicles to seamlessly and securely access the decentralized fog network. Another work on message authentication scheme is introduced in \cite{noh2020distributed} for connected vehicles based on blockchain. It ensures anonymity and decentralization of the broadcasting messages which are exchanged by the connected vehicles. In particular, it enables the vehicles to authenticate the messages in efficient and distributed manner. Here, for authentication, the public key based cryptography and message authentication code techniques are employed. With the help of PBFT and PoW consensus mechanisms, this proposed scheme is able to make a decentralized network. 

\begin{figure} [h]
	\includegraphics[width=\linewidth]{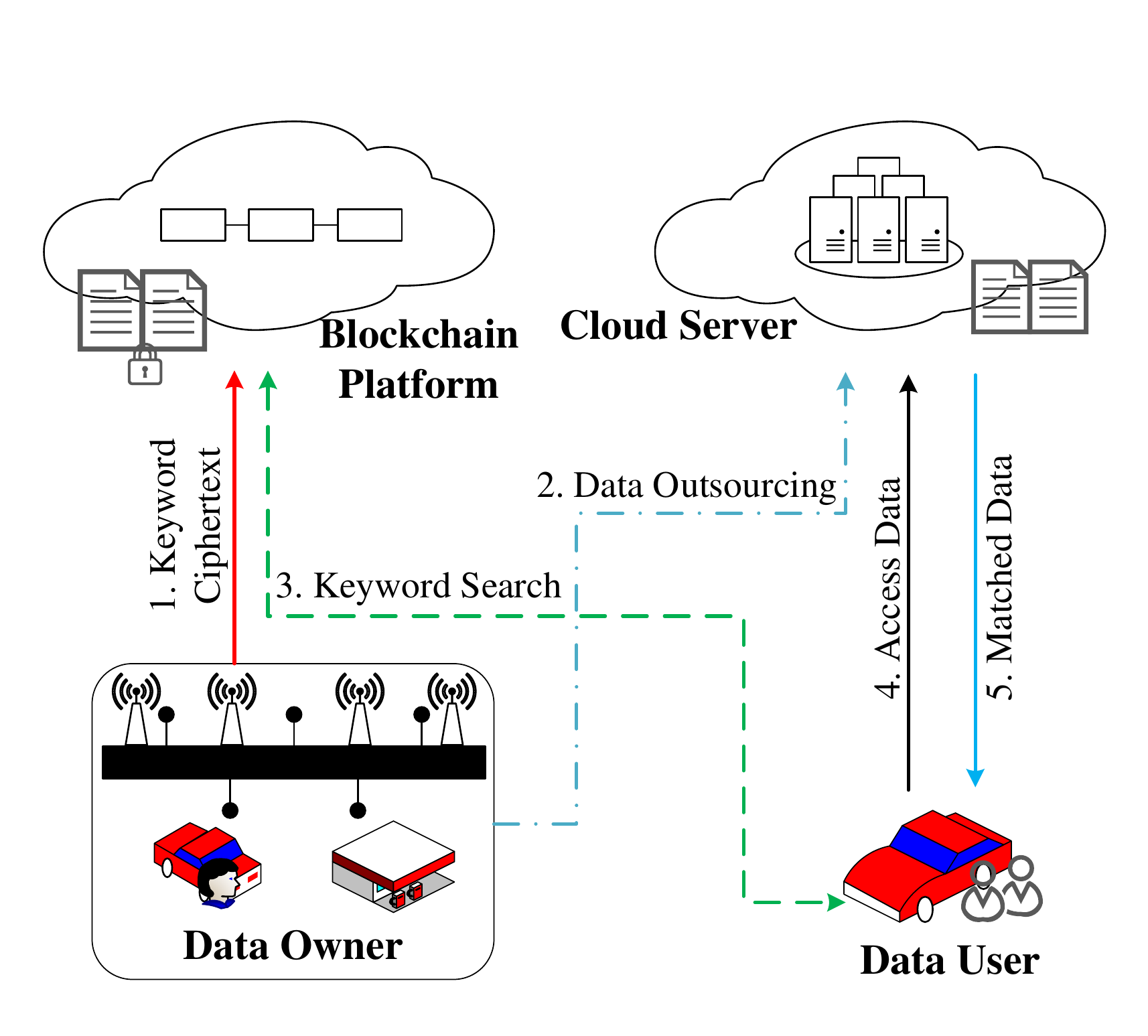}
	 \caption{An overview of the system model of the decentralized searchable cryptographic scheme presented in \cite{chen2019blockchain}}
	 \label{fig: searchablecryptography}
\end{figure}

To realize remote attestation, a security model referred as RASM (Remote Attestation Security Model) is presented in \cite{xu2018remote} for intelligent vehicles in the V2X network, based on privacy-preserving blockchain. This security model is designed to increase the privacy-preserving security while ensuring decentralization, traceability, and non-repudiation. RASM consists of two main steps. In the first step, the identity authentication where the vehicles share their credible identities to the blockchain network as an evidence. In the next step, the vehicles will compute and estimate certain criteria to commit their own decisions. This step also includes the summarization of records to store in the blockchain by some pre-selected nodes. Most importantly, a decentralized searchable encryption scheme for blockchain and cloud-based vehicular social network applications is introduced in \cite{chen2019blockchain} in order to provide efficient keyword searching ability to the users. The searchable encryption is a cryptographic primitive which allows data searching in storage like cloud servers, while ensuring the user data security. To preserve privacy of the users, this scheme also offers forward and backward privacy. Thus, this scheme is referred as BSPEFB. Fig. \ref{fig: searchablecryptography} illustrates the system model and work flow of this proposed BSPEFB. For decentralization, smart contract is adopted in the blockchain to address the single point of attack and check whether the returning result is correct or not.

\begin{figure} [h]
	\includegraphics[width=\linewidth]{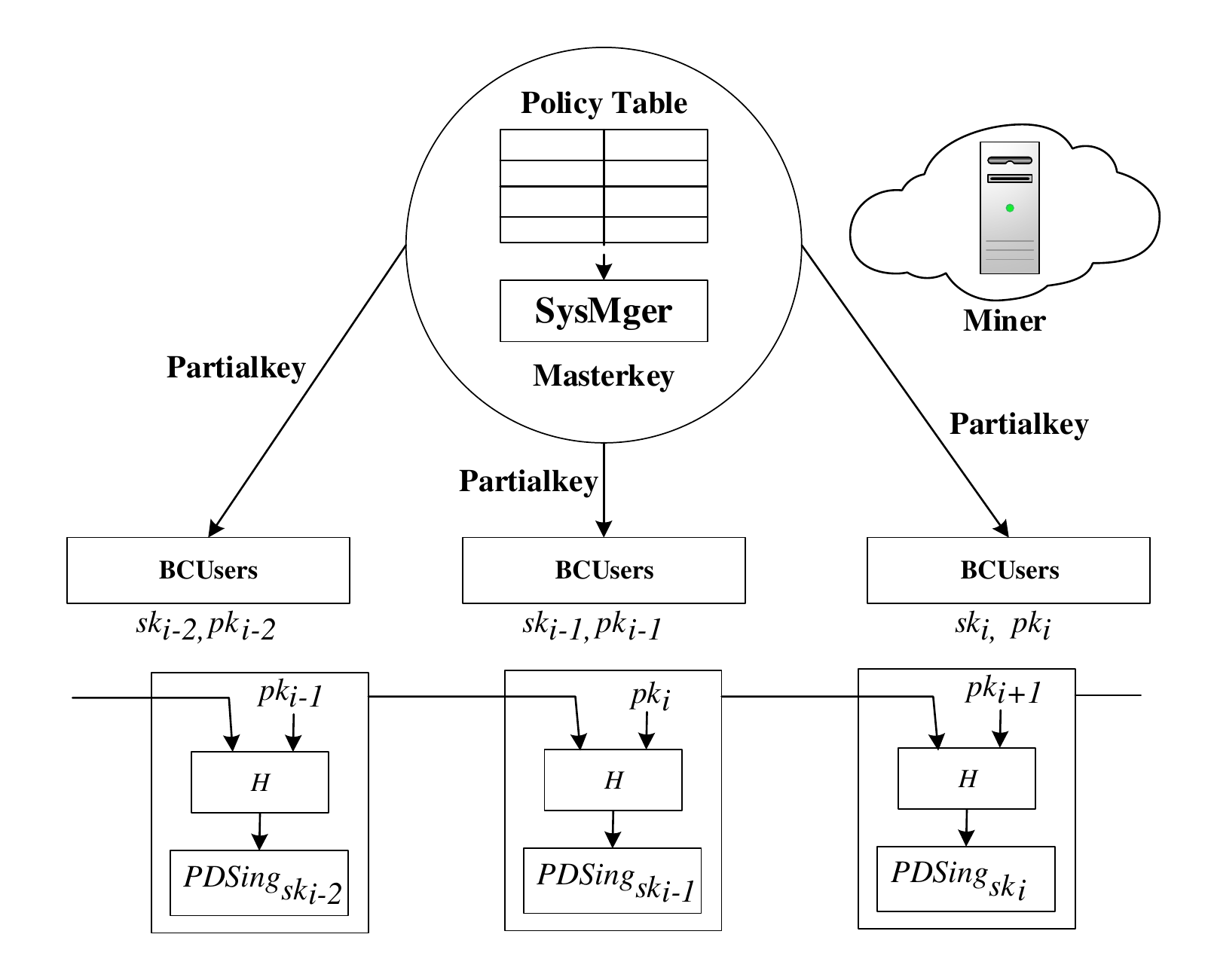}
	 \caption{The system representation of policy-driven signature scheme presented in \cite{mu2019policy} for blockchain-assisted transportation system.}
	 \label{fig: policysignature}
\end{figure}

In order to control the access, a novel and lightweight policy-driven signature scheme referred as PDS is introduced in \cite{mu2019policy} for blockchain-assisted transportation systems. This PDS is particularly developed for private and permissioned blockchain systems since these types of transactions generally have some access rights. A policy set is considered to control the access to ensure that only signers who met the policy set are able to take part in certain transactions, as shown in Fig. \ref{fig: policysignature}. The users in the blockchain network should have signing keys that contain a policy set. Moreover, this scheme replied on attribute-based signature (ABS) and certificateless cryptography. The purpose of this scheme is to offer short signature size and less computation processing time. Compared to traditional ABS schemes, the PDS is more suitable for blockchain requirements due to the adoption of certificateless cryptography which does not require any trusted certificate authority. Lastly, a scheme for location privacy protection of vehicles that use location based services in blockchain-assisted intelligent transportation systems is presented in \cite{luo2019blockchain}. The aim of this proposed scheme is to address the limitations of a popular privacy protection technique named distributed k-anonymity, such as failure to detect malicious vehicles and sensitive location privacy leakage. This scheme considers trustworthiness of vehicles, and based on it, the vehicles cooperate with each other. To keep the records of trustworthiness publicly available for the vehicles, a data structure is introduced for blockchain.

\subsection{Performance}
Incorporating blockchain into IoV may require the capability of handling a large number of data \& transactions within a highly dynamic environment of moving vehicles. This is a current limitation of blockchain technology to accommodate directly into IoV. Hence, because of the dependency of massive data and mobility, the performance metrics of blockchain-enabled IoV network is as important as its security and privacy. These performance matrices include latency, energy consumption, throughput, and scalability.

A few works focus on improving the performance metrics particularly for blockchain-enabled IoV. For example, the authors in \cite{sharma2018energy} address the problem associated with high energy consumption by the vehicles due to the excessive amount of transaction-related messages required for updating the ledger between the vehicles and blockchain network. To save a maximum amount of energy, a distributed clustering approach is introduced where the amount of transactions are controlled optimally. The performance of this proposed approach is discussed in a quantitative manner by numerical analysis which results in this approach saving energy by 40.16\% and decreasing the amount of transactions by 82.06\% compared with the traditional one. Another work in \cite{liu2019deep} presents a performance optimization framework based on the deep reinforcement learning (DRL) technique for blockchain-assisted IoV to address the scalability issue and deal with big data generated by vehicles. The framework focuses on maximizing the transactional throughput while considering three main blockchain properties such as decentralization, security, and latency. In this framework, at first, a quantitative measurement methodology is presented to evaluate the performance of blockchain. Next, the DRL technique is utilized to enhance the performance by selecting the block producers and adjusting the block size \& the interval time according to the dynamic distribution of vehicles.

\subsection{IoV-specific and Optimized Consensus}
The most popular and successful consensus mechanisms are primarily developed for cryptocurrency applications. However, due to the different nature of IoV, there are some challenging issues related to utilizing these mechanisms in IoV, such as block validation, security, reward \& penalty schemes, and energy consumption. 

The IoV-specific and optimized consensus mechanisms are able to increase the applicability of accommodating the blockchain into IoV. As an example, an optimized and security enhanced Delegated Proof-of-Stake (DPoS) consensus mechanism particularly for blockchain-based IoV application is presented in \cite{kang2019toward}. The contribution of this work has two parts. The first part is a reputation-based voting scheme to select miners effectively which is based on multi-weight subjective logic model. The reputation concept aims to prevent collusion between rich stakeholders and candidates. The multi-weight subjective logic model is utilized to compute accurately the reputation of candidate miners. To calculate the reputation of miner candidates, this scheme considers past interactions as well as recommendations. However, the high-reputed candidates will be divided into two categories such as active candidates and standby candidates. On the other hand, the second part is provided incentives to the highly-reputed standby candidates who verify the new blocks after auditing once more. This verification has an aim to reduce the collusion among active candidates. The contract theory is utilized in this second part to model the interactions within the active and standby candidates, where security, delay time, and optimized utilities of both block manager \& verifiers are considered.

In an effort to address the problems which IoV are facing currently due to centralized dependency, such as security, performance, and communication, the authors in \cite{hu2019blockchain} focuses on developing a consensus algorithm for Internet of Vehicles (IoV). At the beginning, the authors modified the current centralized architecture in accordance with the decentralized blockchain architecture by eliminating the central node, dividing all nodes into Internet connected vehicle nodes \& roadside nodes, and making all these nodes equally necessary. Next, the authors adopt the Byzantine consensus algorithm so that all nodes are able to be authenticated without depending on any centralized authority. Moreover, this algorithm also allows the nodes to correctly reach a sufficient consensus in spite of malicious nodes of the network failing or spreading false data to other nodes. Besides, this proposed algorithm utilizes time sequence technique and gossip protocol to enhance the communication and consensus efficiencies. In such a context, an offloading method is presented in \cite{de2019energy} so that the validation process of proof of stake (PoS) can be offloaded to edge or cloud computing platforms. This method is referred as ECbroker which is defined in terms of Satisfiability Modulo Theories (SMT). The offloading of validation process aims to validate more transactions with high throughput and to disburden the processing overhead. The theoretical analysis demonstrates that ECbroker achieves 77.7\% more profit, while consuming 39.2\% less energy  when the vehicles offload their validation processes. With SMT, the ECbroker offers the vehicles to determine whether to validate locally or offload the processes to the edge/cloud. While deciding, the vehicles also considers their computational resources and mobilities. 

\subsection{Incentive Mechanisms}
The IoV based on public and consortium blockchains are usually relied on the contributions of one or more than one entities. Such contributions include new block verification, validation, related data production, and blockchain storage. To provide incentives to the contributors will be the key to expedite blockchain related activities. 

\begin{figure} [h]
	\includegraphics[width=\linewidth]{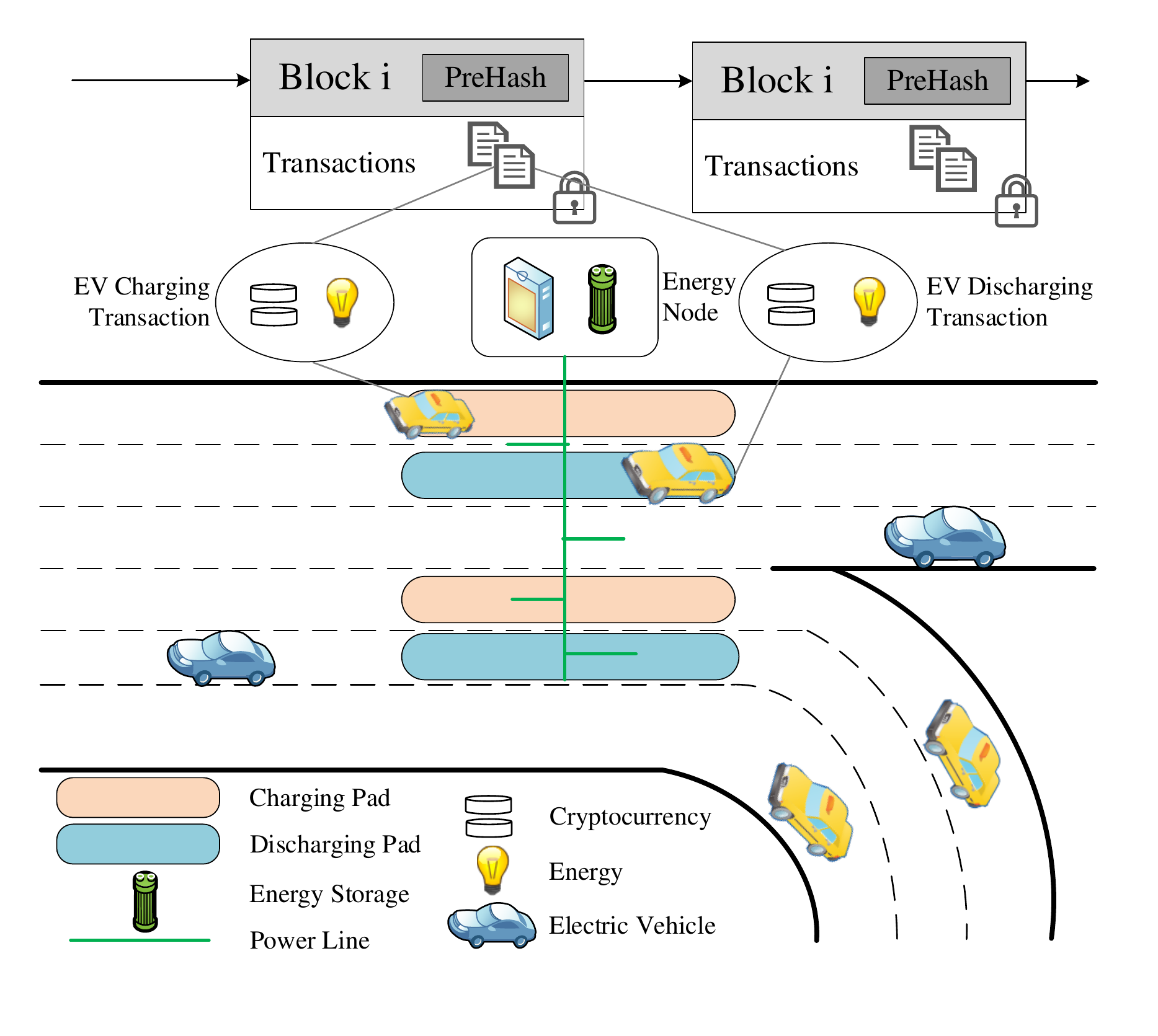}
	 \caption{The network model of incentive mechanism presented in \cite{wang2019bsis}.}
	 \label{fig: networkmodel}
\end{figure}

Some incentive mechanisms are presented which are introduced for IoV application. For example, with the help of permissioned blockchain, an incentive mechanism is introduced in \cite{wang2019bsis} for vehicular energy networks which is referred as BSIS. Fig. \ref{fig: networkmodel} depicts the network model of BSIS. This incentive mechanism is based on the pricing theory to schedule optimally the charging and discharging process of vehicles in order to maximize the utilities of vehicles as well as to acquire the regional energy balance. Moreover, a proof of reputation (PoR) consensus mechanism is also introduced to support the BSIS, where the validators are chosen by their reputations. This reputation technique is employed to enhance the security of the blockchain network.

Meanwhile, another novel incentive scheme is presented in \cite{niyato2019incentivizing} to improve the security of delegated PoS (DPoS) consensus mechanism for blockchain-assisted vehicle data sharing. The purpose of this incentive scheme is to address the potential internal collusion attack while verifying the new blocks. In this attack, the active validators could collude with other false results which are generated maliciously to receive rewards. However, to solve this issue, the new block will be sent to the standby validators for further auditing and verification. Based on contract theory, the incentive mechanism is designed so that the active and standby validators can interact with each other, and the standby validators can get a part of incentives due to their contributions. The numerical results indicate that this incentive scheme is able to optimize the utilities of all validators while enhancing the security of DPoS by protecting it from collusion attack. Lastly, in \cite{ijazreward}, an effort to develop a decentralized reward and penalty mechanism is presented in the context of blockchain and vehicular network which is based on RSU. The aim of this mechanism is to ensure the blockchain-assisted vehicular network more trustworthy. This mechanism considers trust value which is rating of the individual vehicle. The responsibility of RSU is to allocate and manage the trust values. The vehicles in the network receive penalty and reward according to their activities. For instance, once a RSU detects a malicious massage sent by a vehicle, the RSU will produce a negative rating for that particular vehicle. This negative rating ultimately indicates low trust value of that particular vehicle, and it will receive penalty. Moreover, the vehicles having low trust values are not allowed to generate any new messages. By this way, the network will be able to increase the trustworthiness and efficiency among the connected vehicles.

\subsection{Remarks}
In this section, we have presented the challenges associated with integration of blockchain with IoV such as security \& privacy, performance, IoV-specific \& optimized consensus, and incentive mechanisms. Moreover, we have also highlighted the potential solutions to address these four challenges presented in different literatures. These works are summarized in Table \ref{tab: iovintegrationchallenges}.

\begin{table*}[h!]
\centering
    \caption{Summary of literatures on the recent works in blockchain and IoV integration challenges}
    \label{tab: iovintegrationchallenges}
    \begin{tabular}{m{1.1cm}|m{4.1cm}|m{4.2cm}|m{7cm}}
\hline \hline
References & Focused Challenges & Approaches & Outcomes \\
\hline
\cite{lu2018privacy} & - Identity privacy of vehicles through tracking attacks

- Forged messages publishing from internal vehicles & - Anonymous reputation system

- Calculating reputation by historical interactions as well as opinions

- Pseudonym addresses instead of real identities & - The proposed model can build a trust model and also, satisfy the conditional anonymity, transparency, and robustness \\
\hline
\cite{xu2018remote} & - Centralized privacy solution problems & - Remote attestation security scheme & - The proposed model is able to satisfy decentralization, user anonymity, and traceability \\
\hline
\cite{kaur2019blockchain} & - Lack of decentralization feature in modern authentication schemes & - Elliptic curve cryptography (ECC) for mutual authentication & - An authentication and key-exchange scheme having lightweight, scalability, decentralization, and anonymity features \\
\hline
\cite{lei2017blockchain} & - Insecure key management & - A proposed novel network topology

- Secure group broadcast
 & - The proposed framework accomplishes better distributed key management with reduced key transfer time \\
\hline
\cite{chen2019blockchain} & - Lack of decentralized efficient keyword searching schemes & - Searchable encryption

- Smart contract & - The proposed scheme can effectively increase the privacy protection by supporting forward and backward privacy

- The centralized searching technique is replaced by smart contract \\
\hline
\cite{mu2019policy} & - Unlike public, permissioned blockchain users have some constraints & - Authorization based on assigned policy

- Attribute-based signature

- Certificateless cryptography & - This proposed signature scheme possesses short signature size as well as low computational overhead \\
\hline
\cite{noh2020distributed} & - Secure authentication and privacy protection problems & - Message authentication code

- Public-private key pair & - The proposed authentication scheme ensures that it is secure from a number of common attacks

- The scheme has two properties such as decentralization and anonymity \\
\hline
\cite{luo2019blockchain} & - Potentiality of disclosing vehicle location privacy such as location tracings and sensitive information leakage while using location based services & - Dirichlet distribution for trust management

- Storing the trustworthiness by proposed data structure & - This proposed scheme is able to defend trust model attacks, ensure location privacy, and detect malicious vehicles efficiently \\
\hline
\cite{sharma2018energy} & - The energy and number of transactions burdens while updating the distributed ledgers and blockchain-transfer operation & - Controlling the number of transactions optimally by an approach called distributed clustering & - The proposed model consumes considerably less amount of energy and requires less number of transactions compared with the Bitcoin \\
\hline
\cite{liu2019deep} & - The scalability issue of blockchain to manage the massive IoV data & - Deep reinforcement learning & - The proposed framework is capable of maximizing the throughput of transactions while ensuring low latency and decentralization \\
\hline
\cite{kang2019toward} & - Security issues in DPoS consensus mechanism

- Potential collusion among miner candidates and attacked high-stake vehicles while choosing miner candidates by stake-based voting & - Reputation-based miner selection from candidates

- Two step verification and auditing by active and standby miners

- Contract theory & - The proposed security-enhanced mechanism establishes it excellencies in terms of defending internal collusion, high detection rate of compromised candidate vehicles with better reputation scheme relative to existing schemes, and optimized the utilities of all participant miners \\
\hline
\cite{hu2019blockchain} & - Security and authentication concerns in consensus mechanisms & - Byzantine consensus along with gossip protocol and time sequence technique & - The proposed consensus performs better in terms of consensus efficiency, communication security, scalability, and fault tolerance than the traditional mechanism \\
\hline
\cite{de2019energy} & - Incapability of vehicles to participate in competitive PoW and PoS like consensuses to get rewards due to limited resources & - Satisfiability Modulo Theories

- A proposed brokerage method having decision making capability & - The proposed approach offers much higher profit and lesser energy consumption while uploading the mining and validation processes \\
\hline
\cite{wang2019bsis} & - Lack of proper incentive scheme for electric vehicles to contribute in energy delivery & - Pricing and reputation theories & - The proposed scheme not only inspires the vehicles to participate in blockchain network to make a balanced grid, but also enables to maximize he utilities of vehicles \\
\hline
\cite{niyato2019incentivizing} & - Lack of incentive scheme for standby miners & - Contract theory & - The proposed scheme offers improved and secure data sharing \\
\hline
\cite{ijazreward} & - Lack of penalty schemes for generating false messages & - Managing and imposing trust values to vehicles & - This proposed scheme offers more secure and trustworthy network while making restriction from sending false messages \\
\hline
\end{tabular}
\end{table*}

\section{Future Research Opportunities}
Besides the aforesaid remarks and discussions, there are still a number of open challenging issues as well as new research directions which could be considered as future research opportunities. In this section, we will present such opportunities elaborately where we will point out how blockchain along with other state-of-art technologies has further potentially to augment the capabilities of IoV to build the future transportation system.

\subsection{Security of Offloading}
Unlike typical lightweight IoT smart devices which are often resource-constrained, the smart vehicles have enough computational resources such as CPU, memory as well as energy supplies. Although these resources enable the smart vehicles to meet the substantial requirements to participate in computationally expensive new block validation processes, due the mobility nature, the smart vehicles may go to edge computing paradigm to facilitate these processes. Meanwhile, in IoV scenarios, the edge nodes like RSUs are stationary, and are able to offer a convenient way to perform blockchain computations and storages. In \cite{kang2018blockchain, li2019computing, qiao2019blockchain, nkenyereye2020secure}, the authors have presented numerous approaches where they adopted such scenarios. Particularly, the vehicles may offload the validation processes to the edge nodes like RSUs.

However, the edge nodes being utilized could actually be third party owned RSUs which are ultimately not in the trust domains of the vehicles. Hence, it may introduce security issues within blockchain-assisted IoV applications. Specifically, one key issue is that the obtained results of blockchain processing from edge nodes may not be fully trusted and correct. In order to ensure the trustworthiness of the results, verifiability of the correctness of edge processing should be ensured. Hence, a notion named \textit{verifiable computing} is introduced to achieve such goals. This verifiable computing offers the relatively lightweight entities to offload the computational tasks of some functions $F$ on various inputs to other more powerful computing nodes. The powerful computing nodes may be untrusted; thus, they will need to send back the results with proofs so that the lightweight entities will be able to verify that the computations have been performed correctly. Therefore, the design of an efficient verifiable computing scheme will be one of the key scopes for future research in edge computing and blockchain empowered IoV system to validate the correctness of blockchain computational processes done by edge nodes.

\subsection{PKI for Blockchain-Enabled IoV}

For application scenarios in where consortium and private like permissioned blockchains have been employed, the security should be relied on PKI. Here, the PKI refers to an approach which enables the identities of devices, people, and other relevant entities to be affixed with respective public keys. The public keys are transferred over the Internet, and later, used for encryption and digital signature verification purposes. This affixing approach is formed by a method of certificate issuance offered by trusted entities named as certificate authorities (CA). The CA enables anyone to check if an individual public key has been assigned to a specific entity. Thus, each entity has to be connected with the Internet as well as CA in order to attain the PKI. Besides this, the PKI has one another trusted entity called registration authority (RA) which is responsible for maintaining the assurance of validation and correctness of registration processes. 

In the same connection, to deploy secure communications with the public-key encryption techniques, the PKI is an advantageous option for IoV application scenarios. According to Section III, in \cite{javaid2019drivman, cebe2018block4forensic}, such PKI has been employed to support various blockchain-assisted IoV cases.

On the other hand, applying the classic PKI directly into IoV can cause latency, availability, and scalability issues particularly for the vehicles, since the PKI is relied on certificate and registration authorities. Consequently, it is important to develop some novel approaches so that the aforementioned issues can be addressed, and the functionalities of PKI can be fulfilled. Specifically, the authorities should be placed locally so that the vehicles will be able to realize the services with low latency. As such, to develop such approaches, smart contract can be a suitable and timely candidate to be employed. The aim of employing smart contract in PKI is to develop approaches which will be able to resist availability attacks while maintaining low latency communications. At the same time, smart contract will be able to ensure trust if the certificate issuance and verification authorities are placed near to the vehicles. Thus, there is an immense potential to inquire further into the area of PKI for IoV research.

\subsection{Deferentially-Privacy Preserving Solutions}
In a number of IoV cases, it is required to send the data generated by the vehicles to the cloud platform, RSUs, and neighbor vehicles. Also, some vehicles have to share the driving information along with locations to the peers in order to enhance driving experience and to take advantage of location-based services. A number of such IoV applications have been discussed in Section IV. However, certain privacy concerns in this area have arisen recently, as malicious users may find out the real location, location history, live location, and real identity of the vehicles. These aforesaid critical privacy issues could arise mainly due to the following reasons and occasions:
\begin{enumerate} [label=(\roman*)]
\item The RSUs might be maintained by the third-party service providers and they may offer less privacy than the cloud service providers, and the vehicle users might not have control over their own generated data transactions;
\item The offloaded tasks could contain private or sensitive data;
\item While utilizing the data for training the machine learning models as well as communication the model updates;
\item Public storage of blockchain transactions; and
\item After analyzing the movement and communication behaviors of the vehicles as well as the shared data transactions
\end{enumerate}

With the purpose to prevent any information leakage, it is very important to ensure the privacy requirements. In the same connection, in order to deal with privacy concerns, classic blockchain technology offers a privacy-preserving technique to hide the real identities of the vehicles using pseudonyms. Besides this, several cryptographic techniques have been employed as presented in Section V to ensure the privacy of the transactions in blockchain.

Unfortunately, although cryptographic techniques-based approaches are considered as a leading strategy, sometimes such approaches may not be robust. Moreover, it may come with certain limitations like managing cryptographic keys, require significant computational resources, and sacrifice the performance. Meanwhile, it is necessary to explore the state-of-the art privacy-preserving approaches in order to tackle future novel issues that emerged due to having different data structure, architecture, and key techniques of blockchain technology. In this regard, differential privacy which is a privacy-preserving approach could be an option to explore.

In differential privacy, basically a large amount of data like datasets as well as real-time data can be preserved by adding a desirable amount of noise while maintaining a good trade-off between accuracy and privacy. Furthermore, differential privacy also offers location privacy in real-time within the IoV scenarios. This is by doing data perturbation of location as well as identity to ultimately ensure the privacy of the location and vehicles. Usually, this perturbation method is performed in many of differential privacy algorithms. In this method, the amount of noise is computed by employing differential privacy, then this noise will be added to the query data in order to ensure that the data will be secure as well as indistinguishable to the observers. However, this perturbation approach will result on the accuracy of the data which has been reported, and simultaneously, the more perturbed data will tighten the privacy protection. Here, one advantageous point is that the users have the ability to control the privacy level. However, this accuracy and privacy trade-off sometimes may introduce difficulties to certain IoV applications that need accurate reporting of data after incorporating with blockchain. In summary, we believe that differential privacy has a lot of potential to enhance privacy substantially but in different ways.

\subsection{Exploring other Blockchain-Assisted IoV Applications}
We have presented in this paper a number of IoV application scenarios where blockchain has been employed successfully. However, given the benefits of better decentralization, autonomy, and security, we believe blockchain and/or smart contract will be able to play an important role in many other applications as well. In the following, we present diverse types of such applications where blockchain can also be applicable.

\textit{1) Road Safety Applications:} Obtaining the data generated by the vehicles, IoV enables real-time data processing. In case of any hazardous data, the nearby RSUs and vesicles will be responsible for keeping the record and inform to all so that others will be safe from the potential risk by taking necessary steps in advance. In this scenario, the trust will be an important aspect. Thus, we believe that after integrating blockchain with road safety applications, which could be considered as future works, the vehicles will be able to trust the blockchain network established by RSUs and other peer vehicles.

\textit{2) Vehicular Big Data Auction:} In the past few years, numerous auction theories have been introduced and applied successfully which have common aim to perform auctions of different services and products between the two parties such as sellers and buyers. In the same connection, the auction theories have also been adopted in vehicular big data market in order to carry out auctions. The features of such auctions even having predefined and trusted centralized authorities are already acceptable to the traders, but on the other hand, several concerns have been introduced. Such concerns are mainly due to processing cost and possibility of both insider and outsider threats. Thus, smart contract in conjunction with blockchain is will be one of suitable solutions to address the concerns.

\textit{3) Content Caching:} With the exponential increasing of data traffic, utilizing storage resources of RSUs and vehicles to server as a supplement of the cloud for the purpose of caching popular contents have become gradually popular. In particular, the applications and services where the data consumptions are high compared to others have been performed with the support of content caching. This content caching enables less dependency on backhaul connection while ensuring low latency communications as well as better user experience. Hence, how to utilize blockchain technology for efficient, fast, and economically viable content caching is very important. In IoV scenarios, blockchain and smart contract would contribute in two ways. The first is the exploitation of blockchain structure by keeping only the lightweight block headers of the cached contents, which would able to enforce trustfulness among the vehicles and cache nodes significantly faster than the classic caching system. The another is developing economic incentive-friendly smart contract for the cache contributors, which would lead to encourage the contributors.

\subsection{Distribution of Economic Profits and Incentives}
Blockchain technology has been developed for the application cases which have stable and stationary network connections. On the other hand, in IoV ecosystem, the vehicles are mobile, move from one place to another very frequently, and try to make connection with peers by wirelessly as long as communication conditions will be allowed. As a result, making blockchain suitable for such IoV scenarios which may often suffer from poor and unstable wireless connection is one of the challenging tasks. Furthermore, with the development of blockchain, the spectrum, storage, and computational resources can be shared among the peers as discussed in section IV. Blockchain has made easier to establish peer to peer such resource sharing. However, recently a number of research questions have been arisen which could be future research to be considered. These are discussed shortly as follows.

\begin{enumerate} [label=(\roman*)]
\item How to aggregate the resources to utilize collaboratively by vehicles, \& edge nodes, and incorporate the aggregated resources in blockchain ecosystem;

\item How to split and distribute the profits among the edge devices, service providers, vehicles, brokers, and the participants who have contributions for aggregation so that all entities will receive the profits rightly according to their contributions;

\item How to realize the different resources as quantified values, use pricing theories, and optimize the profits of all parties so that they are willing to participate;

\item How to develop suitable and practical approaches in order to make connections with surrounding nodes while keeping long-term stable link;

\item How to ensure and maintain the trustworthiness of entities by using blockchain and/or smart contract while trading resources;

\item How to detect malicious and dishonest entities who may have intention to repudiate the transactions and deny giving the incentives or agreed resources;

\item How to quantify the profits and incentives properly for the contributors who have dropped out and went offline because of the wireless connectivity;
\end{enumerate}

\section{Conclusion}
In this paper, we have presented a survey on the integration of blockchain and IoV from the ITS perspective, which is becoming a key technique which utilizes decentralized management and stronger security solutions. We have started our study with some preliminary background, including brief discussions on blockchain technology, edge computing, Intelligent Transportation System, and Internet of Vehicles. We have also discussed the motivations of this study by identifying the challenges associated with IoV and also, the realizations of decentralization, high immutability, availability, and trust. Then, we have conducted a systematic and comprehensive investigation on blockchain in IoV paradigm, in terms of application scenarios, how other techniques have been incorporated along with blockchain, and functionalities. We have further highlighted four key challenges after utilizing blockchain in IoV, and also, presented related research works addressing these challenges, including security \& privacy, performance, and IoV-specific \& optimized consensus as well as incentive mechanisms. Next, we have discussed a number of blockchain empowered architecture and frameworks in details. Finally, we have tried to explore future research opportunities in this domain briefly. To sum up, it is expected that the blockchain technology along with IoV will significantly include new functionalities to transportation systems. We hope this paper will serve as basis to the research community to move forward.



\ifCLASSOPTIONcaptionsoff
  \newpage
\fi

\bibliographystyle{IEEEtran}
\bibliography{bibliography.bib}

\begin{IEEEbiography}[{\includegraphics[width=1in,height=1.25in,clip,keepaspectratio]{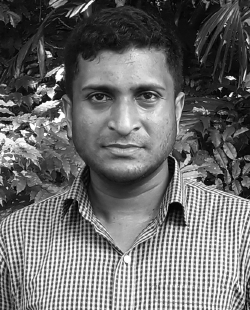}}]
{Muhammad Baqer Mollah} is currently working as a Research Associate in the School of Computer Science and Engineering at Nanyang Technological University (NTU), Singapore. Before joining NTU, he was working at Singapore University of Technology and Design (SUTD). He is currently involved in research works on AI and Blockchain applications to the cyber-physical systems (e.g., smart grid, industry, transportation). His research interests are mainly focused on advanced communication, security, and resource allocation techniques for future wireless networks and cyber-physical systems. He has a M.Sc. in Computer Science and B.Sc. in Electrical \& Electronic Engineering from Jahangirnagar University, Dhaka and International Islamic University Chittagong, Bangladesh, respectively.
\end{IEEEbiography}

\begin{IEEEbiography}[{\includegraphics[width=1in,height=1.25in,clip,keepaspectratio]{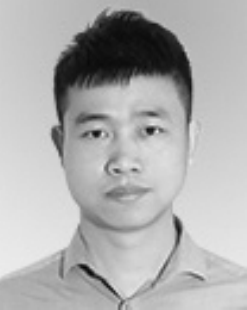}}]
{Jun Zhao} is an Assistant Professor in the School of Computer Science and Engineering at Nanyang Technological University (NTU). He received a PhD degree in Electrical and Computer Engineering from Carnegie Mellon University (CMU) in the USA and a Bachelor’s degree from Shanghai Jiao Tong University in China. Before joining NTU as a faculty member, he was a postdoctoral research fellow at NTU with Prof. Xiaokui Xiao. Prior to that, he was an Arizona Computing PostDoc Best Practices Fellow at Arizona State University working with Prof. Junshan Zhang therein and Prof. Vincent Poor at Princeton University. During the PhD training at CMU, he was advised by Prof. Virgil Gligor and Prof. Osman Yagan, while also collaborating with Prof. Adrian Perrig (now at ETH Zurich). His research interests include security/privacy (e.g., blockchains), wireless communications (eg., 5G, Beyond 5G/6G), and energy system (smart grid, Energy Internet, data center). He has over a dozen journal articles published in IEEE/ACM Transactions as well as over thirty conference/workshop papers.
\end{IEEEbiography}

\begin{IEEEbiography}[{\includegraphics[width=1in,height=1.25in,clip,keepaspectratio]{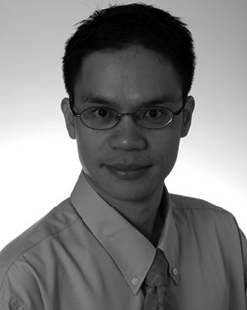}}]
{Dusit Niyato} is a Professor in the School of Computer Science and Engineering at Nanyang Technological University (NTU). He received B.Eng. from King Mongkut’s Institute of Technology Ladkrabang (KMITL), Thailand in 1999 and Ph.D. in Electrical and Computer Engineering from the University of Manitoba, Canada in 2008. He is a Fellow of IEEE. He received several best paper awards from well-known conferences such as IEEE ICC and IEEE WCNC. He is currently an editor of IEEE Transactions on Communications and IEEE Transactions on Vehicular Technology, and a senior editor of IEEE Wireless Communications Letter. His research interests are in the area of wireless communications and networks, game theory, smart grid, edge computing, blockchain technology, and Internet of Things (IoT).
\end{IEEEbiography}

\begin{IEEEbiography}[{\includegraphics[width=1in,height=1.25in,clip,keepaspectratio]{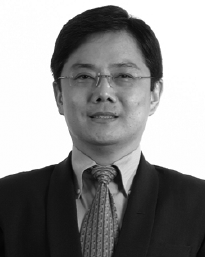}}]
{Yong Liang Guan} is currently a Professor with the School of Electrical and Electronic Engineering, Nanyang Technological University, Singapore. He received the Ph.D. degree from the Imperial College of London, U.K., and the B.Eng. degree (Hons.) from the National University of Singapore. His research interests broadly include modulation, coding and signal processing for communication systems, and data storage systems. He is an Associate Editor of the IEEE TRANSACTIONS ON VEHICULAR TECHNOLOGY.
\end{IEEEbiography}

\begin{IEEEbiography}[{\includegraphics[width=1in,height=1.25in,clip,keepaspectratio]{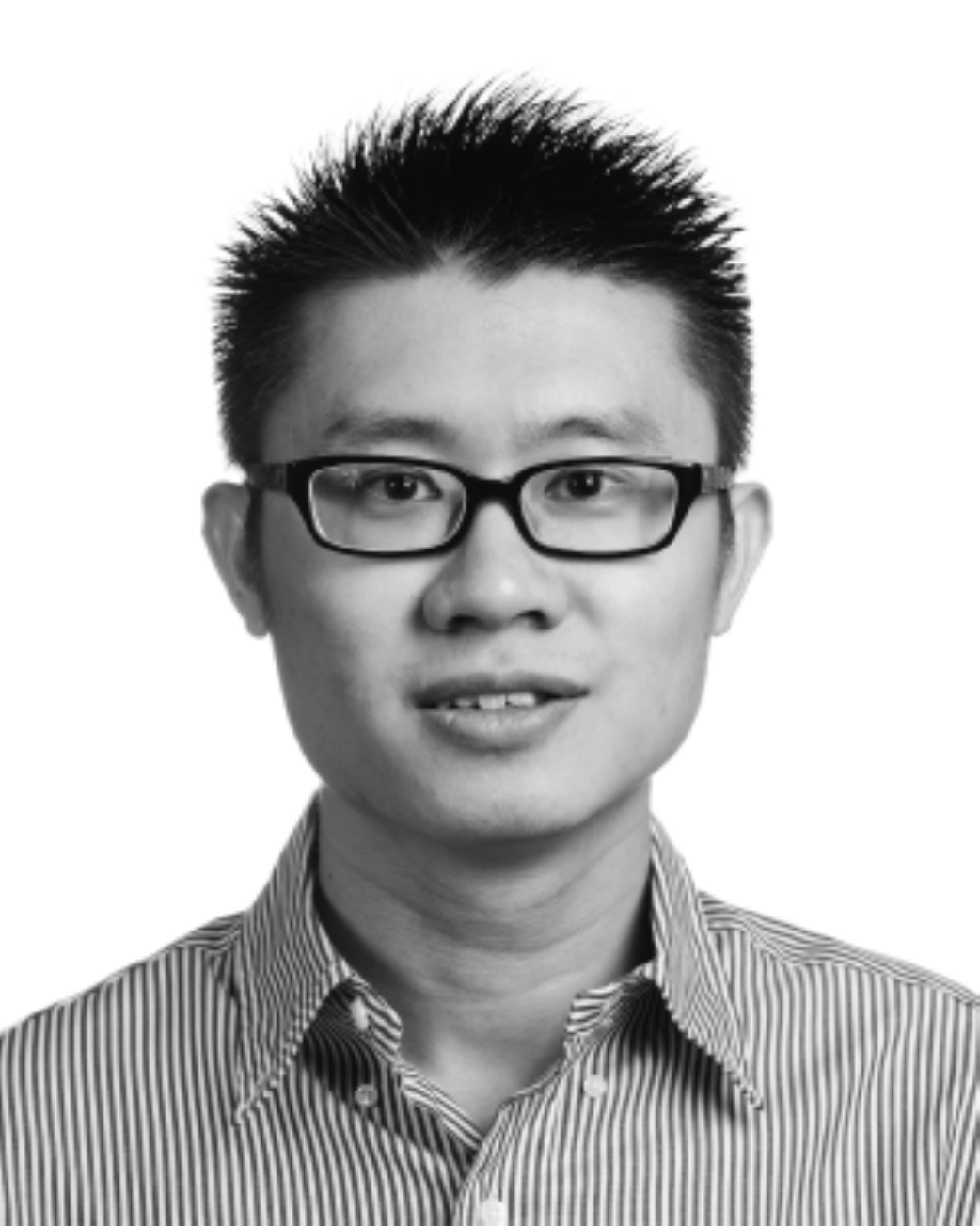}}]
{Chau Yuen} is an Associate Professor with Engineering  Product  Development (EPD) Pillar at Singapore University of Technology and Design (SUTD). He was a PostDoctoral Fellow at Lucent Technologies Bell Labs, Murray Hill, NJ, USA, in 2005. He was a Visiting Assistant Professor at The Hong Kong Polytechnic University in 2008. From 2006 to 2010, he was a Senior Research Engineer at the Institute for Infocomm Research (I2R), Agency for Science, Technology and Research (A*STAR), Singapore where he was involved in an industrial project on developing an 802.11n Wireless LAN system, and participated actively in 3Gpp Long Term Evolution (LTE) and LTE-Advanced (LTE-A) Standardization. He received the B.Eng. and Ph.D. degrees from Nanyang Technological University, Singapore, in 2000 and 2004, respectively. He is a recipient of the Lee Kuan Yew Gold Medal, the Institution of Electrical Engineers Book Prize, the Institute of Engineering of Singapore Gold Medal, the Merck Sharp \& Dohme Gold Medal, and twice the recipient of the Hewlett Packard Prize. He received the IEEE Asia-Pacific Outstanding Young Researcher Award in 2012. He serves as an Editor for the IEEE TRANSACTION ON COMMUNICATIONS and the IEEE TRANSACTIONS ON VEHICULAR TECHNOLOGY and was awarded the Top Associate Editor from 2009 to 2015.
\end{IEEEbiography}

\begin{IEEEbiography}[{\includegraphics[width=1in,height=1.25in,clip,keepaspectratio]{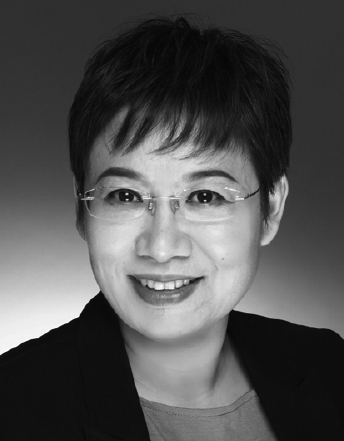}}]
{Sumei Sun} is currently Head of the Communications and Networks Cluster at the Institute for Infocomm Research (I2R), Agency for Science, Technology and Research (A*STAR), Singapore. She is also a Professor with the Infocomm Technology Cluster, Singapore Institute of Technology (SIT), Singapore. She received the B.Sc. degree from Peking University, Beijing, China, the M.Eng. degree from Nanyang Technological University, Singapore, and the Ph.D. degree from the National University of Singapore, Singapore. Her current research interests are in Industrial Internet of Things and  next-generation machine-type communications. She is a Distinguished Speaker of the IEEE Vehicular Technology Society 2018-2021, Vice Director of IEEE Communications Society Asia Pacific Board, and Chapter Coordinator of Asia Pacific Region in the IEEE Vehicular Technologies Society. She is currently a Editor-in-Chief of IEEE Open Journal of Vehicular Technology.
\end{IEEEbiography}

\begin{IEEEbiography}[{\includegraphics[width=1in,height=1.25in,clip,keepaspectratio]{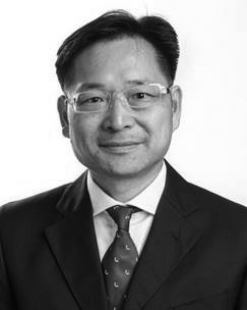}}]
{Kwok-Yan Lam} is currently a Professor at School of Computer Science and Engineering, Nanyang Technological University. Prior to joining NTU, he has been a Professor of the Tsinghua University, PR China (2002-2010) and a faculty member of the National University of Singapore and the University of London since 1990. He was a visiting scientist at the Isaac Newton Institute of the Cambridge University and a visiting professor at the European Institute for Systems Security. In 1998, he received the Singapore Foundation Award from the Japanese Chamber of Commerce and Industry in recognition of his R\&D achievement in Information Security in Singapore. He received his B.Sc. (First Class Hons.) in computer science from the University of London in 1987 and his Ph.D. from the University of Cambridge in 1990. His research interests include Distributed Systems, IoT Security Infrastructure, Distributed Protocols for Blockchain, Biometric Cryptography, Homeland Security, and Cybersecurity.
\end{IEEEbiography}

\begin{IEEEbiography}[{\includegraphics[width=1in,height=1.25in,clip,keepaspectratio]{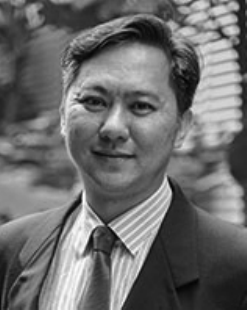}}]
{Leong Hai Koh} is currently a Senior Scientist at the Energy Research Institute @NTU (ERIAN), Nanyang Technological University, Singapore. He received the B.Eng. degree (Hons.) and the Ph.D. degree in electrical engineering from Nanyang Technological University (NTU). His current research interests include smart grid, energy information and management system, hybrid AC/DC microgrid, renewable energy and integration, and power system modeling and simulation.
\end{IEEEbiography}
\end{document}